\documentclass[preprint]{aastex62}

\usepackage{amsmath}
\usepackage{amsbsy}

\defcitealias{Higginson2017b}{Paper~I}

\newcommand{\Alfven}{Alfv\'{e}n\ }
\newcommand{\Alfvenic}{Alfv\'{e}nic\ }
\newcommand{\Alfvenicity}{Alfv\'{e}nicity\ }

\shorttitle{Structured Slow Solar Wind Variability}
\shortauthors{Higginson \& Lynch}

\begin{document}


\title{Structured Slow Solar Wind Variability:\\Streamer-Blob Flux Ropes and Torsional Alfv\'{e}n Waves}


\author{A.~K.~Higginson}
\affiliation{Department of Climate and Space Sciences and
Engineering, University of Michigan, Ann Arbor, Michigan 48109,
USA}
\author{B.~J.~Lynch} 
\affiliation{Space Sciences Laboratory, University of
California--Berkeley, Berkeley, CA 94720, USA}

%

\begin{abstract}

The slow solar wind exhibits strong variability on timescales from minutes to days, likely related to magnetic reconnection processes in the extended solar corona. \citet{Higginson2017b} presented a numerical magnetohydrodynamic simulation which showed interchange magnetic reconnection is ubiquitous and most likely responsible for releasing much of the slow solar wind, in particular along topological features known as the Separatrix-Web (S-Web). Here, we continue our analysis, focusing on two specific aspects of structured slow solar wind variability. The first type is present in the slow solar wind found near the heliospheric current sheet, and the second we predict should be present everywhere S-Web slow solar wind is observed. For the first type, we examine the evolution of three-dimensional magnetic flux ropes formed at the top of the helmet streamer belt by reconnection in the heliospheric current sheet (HCS). For the second, we examine the simulated remote and in situ signatures of the large-scale torsional \Alfven wave (TAW) which propagates along an S-Web arc to high latitudes. We describe the similarities and differences between the reconnection-generated flux ropes in the HCS, which resemble the well-known "streamer blob'' observations, and the similarly structured TAW. We discuss the implications of our results for the complexity of the HCS and surrounding plasma sheet, and the potential for particle acceleration, as well as the interchange reconnection scenarios which may generate TAWs in the solar corona. We discuss predictions from our simulation results for the dynamic slow solar wind in the extended corona and inner heliosphere.

\end{abstract}


\section{Introduction}
\label{sxn_intro}

The global magnetic structure of the solar corona directly
determines the structure of the solar wind outflow
\citep[e.g.][]{Zirker1977,Axford1999, Antiochos2007, Antiochos2011,
Cranmer2012}. Decades of in situ observations have shown that the
heliospheric structure and solar wind properties reflect the coronal
magnetic structure of its origin \citep{Zurbuchen2007}. The high-speed
solar wind that originates from coronal holes is relatively fast,
cool, and homogeneous \citep{Geiss1995, McComas2002}, and the
low-speed solar wind originating from or near large-scale closed
flux systems (i.e. coronal streamers) is hotter, denser, and exhibits
considerably more variability in its in situ plasma, field, and
composition measurements \citep{Gosling1997, Zurbuchen2000,
Zurbuchen2002, Kepko2016}.
While a stationary, steady-state slow speed wind can be obtained
in open flux regions at the boundaries of helmet streamers due to the large expansion factors \citep{Arge2000,
Cranmer2012, WangYM2012}, research increasingly suggests that it
is composition which categorizes the solar wind, rather than speed
\citep[e.g.][]{Stakhiv2015, Stakhiv2016}. Therefore, reproducing the
variability in the slow solar wind plasma composition during the
formation of solar wind has become essential. Dynamic
magnetic reconnection scenarios for the release
of coronal plasma from the closed magnetic field into the solar wind must be invoked
\citep[e.g.][]{WangYM2000, Antiochos2011, Higginson2017a, Uritsky2017}.

Evidence for the importance of magnetic reconnection during the formation of the solar wind exists in both 
remote and in situ observations.
High quality, high cadence white-light coronagraph data and recent
heliospheric imaging data have shown that the slow solar wind
originating near the coronal helmet streamer belt includes an
intermittent and highly structured outflow of intensity enhancements
across spatial scales, the largest of which are known as ``streamer
blobs,'' which move outwards into the heliosphere with the bulk
outflow of the solar wind \citep{Sheeley1997, Sheeley1999, Sheeley2007,
Sheeley2009, Song2009, Rouillard2010a, Rouillard2010b, Viall2015, Sanchez-Diaz2017}.
At the same time, in situ measurements farther out in the heliosphere also find 
small-scale magnetic flux ropes in the slow solar wind.
\citet{Cartwright2010} described the initial \citet{Moldwin2000}
observations as ``[having] bipolar field rotations coincident with
a core field enhancement and were on the order of tens of minutes
duration and displayed the signature of a force-free, symmetric
magnetic flux rope.'' There has been considerable progress in the
characterization of plasma and field properties of these events,
including large statistical surveys \citep{Feng2007, Feng2008,
Cartwright2008, Cartwright2010, Kilpua2009b, Foullon2011, Janvier2014a,
Janvier2014b, Feng2015a,Feng2015b, Yu2014, Yu2016}.
Many of these researchers \citep[e.g.][and others]{Janvier2014b,Yu2014}
have discussed the similarities and differences between small-scale
flux ropes and the larger interplanetary coronal mass ejections
(ICMEs) which contain a coherent in situ flux rope structure, called
magnetic clouds \citep[][]{Burlaga1981, Marubashi1986, Burlaga1988,
Lepping1990}.
Considerable progress has also been made toward linking these small
flux rope in situ signatures, which are almost always observed in
the slow solar wind associated with the heliospheric plasma sheet,
to their source region in the solar corona. For example,
\citet{Kilpua2009b} trace a number of their small-scale flux rope
events back to their origin at the potential field source surface
(PFSS) sector boundaries (i.e. the helmet streamer belt), and
the detailed analyses of \citet{Rouillard2011} directly image the
coronal and heliospheric propagation of several in situ small-scale
flux rope transients via running-difference tracks of their
corresponding coronal streamer blobs in the STEREO Heliospheric
Imager data.

The studies above all focus on magnetic flux ropes as the main source 
of magnetic field variability, but 
there is a second type of variability which can be linked back to solar wind formation.
A number of researchers have examined periods of in situ solar wind
data that have some or all of the signatures of small-scale flux
rope events but are actually consistent with large-scale torsional \Alfven
waves (TAWs) instead \citep[e.g.][]{Gosling2010taw, Marubashi2010taw,
Cartwright2010, Yu2014}. The schematic of the coherent magnetic
field rotation of a propagating torsional \Alfven wave presented
by \citet{Marubashi2010taw} illustrates just how similar to a twisted
flux rope these signatures may be.
The main difference between the two is the direction of the core
field. While the core field of a traditional flux rope is transverse
to the direction of propagation (at least at formation), that of a
torsional \Alfven wave is along the radial solar wind magnetic
field, allowing the twist component of the flux rope-like structure
to propagate along the core magnetic field as a wave.
One key discriminator between this type of propagating ``pseudo-flux
rope'' wave and a true flux rope is the Alfv\'{e}nicity of the
period in question.
\citet{Yu2014}, and others, have argued if the structured, rotation-like
$\delta \boldsymbol{B}$ fluctuations are highly correlated with
$\delta \boldsymbol{V}$ during the event interval, then the period
of interest should probably be considered a propagating \Alfvenic
disturbance rather than a small-scale flux rope.
As we will discuss herein, it is extremely likely that \emph{both}
these torsional \Alfven waves and the traditional small-scale flux
rope transients are generated by coronal magnetic reconnection processes.

{In this paper we extend the analysis of our numerical magnetohydrodynamic 
(MHD) simulation of reconnection and evolution at the Separatrix-web \citep[S-web;][]{Antiochos2011} 
described by \citet{Higginson2017b}, hereafter referred to as \citetalias{Higginson2017b},
by focusing on the two aspects of \emph{structured} slow solar wind variability described
above, flux ropes within the heliospheric current sheet (HCS) and torsional Alfv\'{e}n waves, both of which are present in our simulation.}
The paper is organized as follows. 
In section~\ref{sxn_setup}, we briefly recap the
\citetalias{Higginson2017b} MHD simulation set up and our previous
results.
In section~\ref{sxn_sim}, {we present synthetic remote and in situ signatures 
of our simulated HCS flux ropes and discuss the relevant properties of the 
streamer blob and small-scale flux rope observations.}
We also examine the large-scale TAW which results from our boundary
driving and propagates along the S-Web arc to high latitudes, and
discuss its similarities to in situ measurements.
For both of these types of structured variability, we present visualizations
of magnetic field line connectivity and evolution from a set of
stationary observers in the heliosphere, akin to orbiting spacecraft.
In section~\ref{sxn_disc}, we discuss our results and their
implications for the upcoming Parker Solar Probe and Solar Orbiter
missions, and in section~\ref{sxn_conc}, we present our conclusions.

\section{Numerical Methods}
\label{sxn_setup}

\begin{figure*}
	\includegraphics[width=\textwidth]{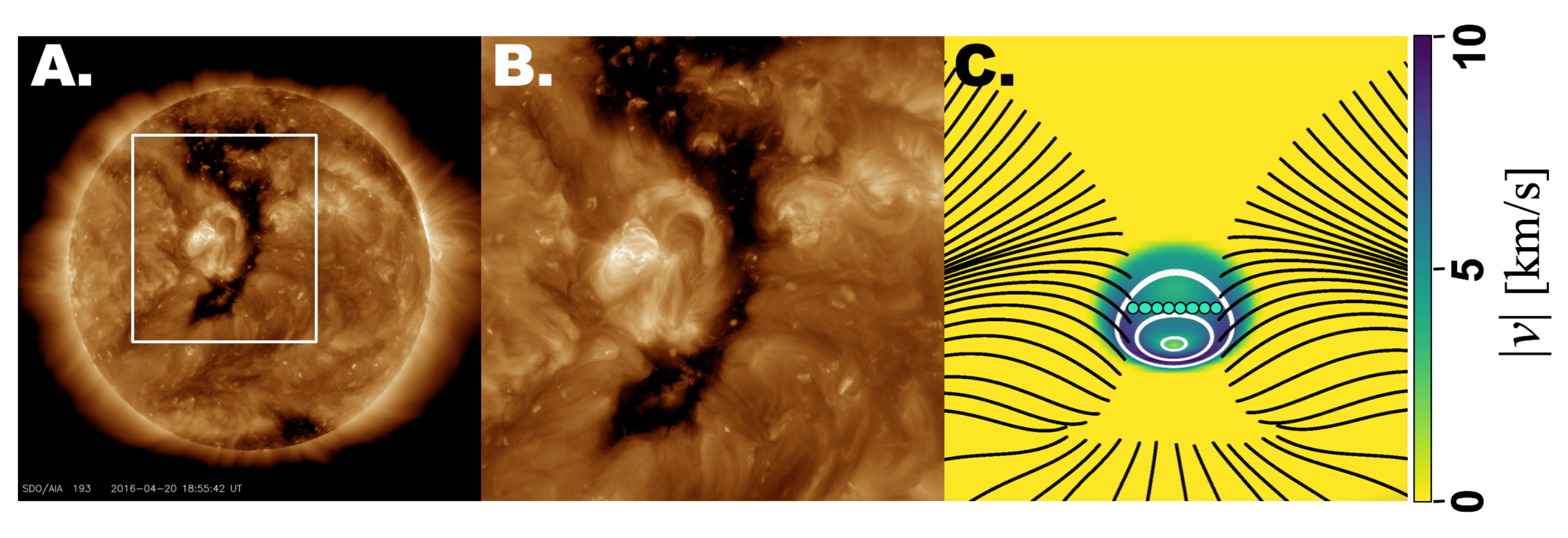}
	\caption{A. \textit{Solar Dynamics Observatory} EUV image of the Sun. 
	The coronal hole appears dark relative to the surrounding corona. B. 
	Zoom-in of white box from A. C. Simulated S-Web corridor. Black 
	magnetic field lines outline the boundary of the polar and low-latitude
	open regions and the connecting open-field corridor on the surface. 
	Cyan dots cross the corridor. Yellow/green/purple shading indicates
	the driving velocity magnitude applied in the simulations; white velocity
	streamlines on the surface show the rotational motion. (Reproduced from \citetalias{Higginson2017b}.)}
	\label{fig0}
\end{figure*}

The numerical simulation presented in \citetalias{Higginson2017b}
is run using the Adaptively Refined MHD Solver \citep[ARMS;][]{DeVore2008}.
ARMS solves the equations of ideal MHD in 3D spherical coordinates
using a finite-volume, flux-corrected transport scheme \citep{DeVore1991}.
Our initial magnetic field configuration, given in \citet{Antiochos2011},
consists of an elongated coronal-hole corridor extending southwards
from a polar coronal hole, like the one shown in Figures \ref{fig0}A and \ref{fig0}B. 
The black magnetic field lines in Figure \ref{fig0}C show the outline of the simulated coronal hole from \citetalias{Higginson2017b} 
on the solar surface. 


The computational domain is defined as $r \in [1R_\odot, 30R_\odot]$,
$\theta \in [11.25^\circ,168.75^\circ]$, where $\theta$ is measured
from the north pole, and $\phi \in [-90^\circ,90^\circ]$. We use
logarithmic grid spacing in the radial direction and linear spacing
in $\theta$ and $\phi$. The grid is constructed using the PARAMESH
package \citep{MacNeice2000} from $8 \times 7 \times 8$ root blocks
with 5 additional levels of static grid refinement specified over
the coronal-hole corridor and resulting S-Web arc. The effective
grid resolution is thus $1024 \times 896 \times 1024$ in the highest
refinement regions.
While there is no explicit physical resistivity included in ideal
MHD, we note that necessary and stabilizing diffusion terms are
inherent in the numerical scheme allowing for an effective numerical
resistivity when the magnetic field gradients are compressed
to the scale of the grid.
The $(r, \theta, \phi)$ boundary conditions and our implementation
of the \citet{Parker1958} isothermal solar wind background are
described by \citet{Masson2013}, \citet{Karpen2017}, and
\citet{Higginson2017a}. The solar wind is initialized with a $T=1$~MK
isothermal atmosphere and the density at the lower radial boundary
is set to $\rho_\odot/m_p = 3.6 \times 10^9$~cm$^{-3}$.

\citetalias{Higginson2017b} described the evolution and dynamics
of interchange reconnection that takes place along a high-latitude
S-Web arc formed by a narrow coronal-hole corridor after it was
perturbed by a supergranular-like, flux-preserving rotational flow. The colored contours 
in Figure \ref{fig0}C show the magnitude of the rotational flow imposed across the coronal-hole corridor 
and the white streamlines show the direction of the flow. This flow lies in the $\theta$,$\phi$ plane only and is 
constructed so as to preserve the normal component of the magnetic field 
during the evolution. In order to satisfy 
\begin{eqnarray}
\label{FlowInductionEqn}
\frac{\partial B_r}{\partial t} = - \nabla_\perp \cdot \left( \mathbf{v_{\perp}} B_r \right) = 0,
\end{eqnarray}
\noindent
we choose $\bf{v_\perp}$ to be equal to the curl of a radial vector, 
\begin{eqnarray}
\label{CurlVector}
\bf{v_\perp} = \nabla_\perp \times \left( \psi, 0, 0 \right).
\end{eqnarray}
\noindent We define $\psi$ as a function of $\theta$, $\phi$, and $t$, 
\begin{eqnarray}
\label{psi}
\psi \left( \theta, \phi, t \right) \equiv V_0 f(t) g(\xi) h(\beta),
\end{eqnarray}
\noindent
where 
\begin{eqnarray}
\label{ComponentsOfPsi}
f(t) &=& \frac{1}{2} \left[ 1 - \cos \left( 2 \pi k \frac{t-t_0}{t_2 - t_1} \right) \right], \\
g(\xi) &=& \frac{(m + l + 1)}{(l+1)} \left[1-\xi^{2(l+1)} \right] - \left[ 1 - \xi^{2(m+l+1)}\right],
\\
h(\beta) &=& \frac{1}{2}\beta^2.
\end{eqnarray}

\noindent In the equations above, the parameters $k$, $t_0$, $t_1$, $t_2$ are set 
to ramp up the flow to maximum velocity from zero and then from that 
velocity back to zero. This ensures that all disturbances are smooth. 
The equation for $g(\xi)$ defines an annulus in spatial coordinate 
$\xi$, where
\begin{eqnarray}
\xi^{2} \equiv 4 \left( \frac{\theta - \theta_{0}}{\theta_{2} - \theta_{1}} \right) ^{2} + 4 
\left( \frac{\phi - \phi_{0}}{\phi_{2} - \phi_{1}} \right) ^{2}.
\end{eqnarray} 

\noindent The location of the flow annulus is determined by the limits $\theta_1$, $\theta_2$, 
and $\phi_1$, $\phi_2$, with coordinate $\left( \theta_0,\phi_0 \right)$ representing the center. 
The thickness and radial velocity profile of the flow annulus are defined by $m$ 
and $l$. We set $m = l = 1$ to yield a thick annulus with a velocity 
peak at the center of the annulus.  In the equation for $h(\beta)$, 
$\beta$ is the magnetic field coordinate between minimum and maximum 
strengths, i.e., 
\begin{eqnarray}
\beta \equiv \max \left( \min \left( B_{r}, B_{2} \right) , B_{1}\right).
\end{eqnarray}
where we chose $B_{1}$ and $B_{2} $ so that $\beta = B_{r}$ everywhere in our flow region.  

\begin{figure*}
	\includegraphics[width=0.5\textwidth]{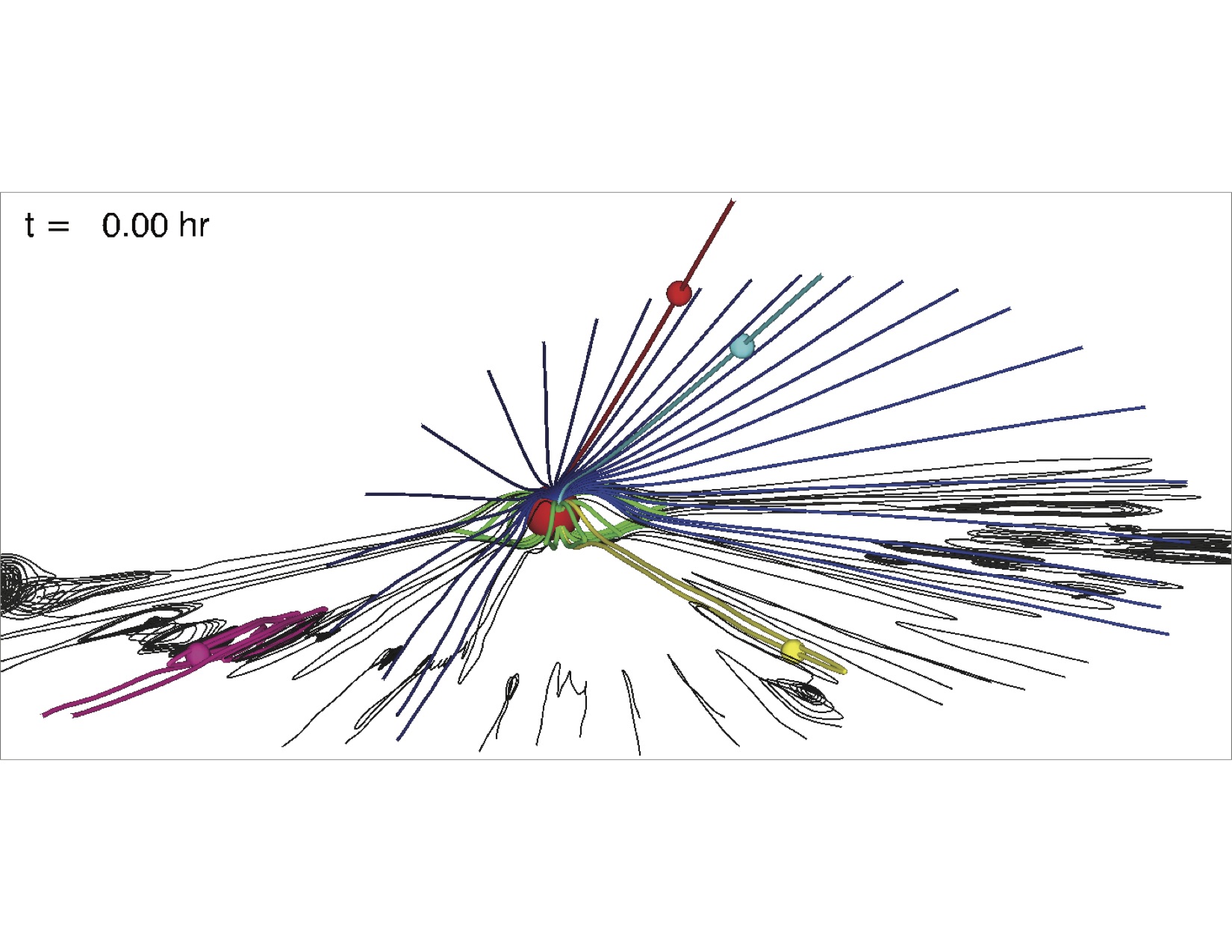}
	\includegraphics[width=0.5\textwidth]{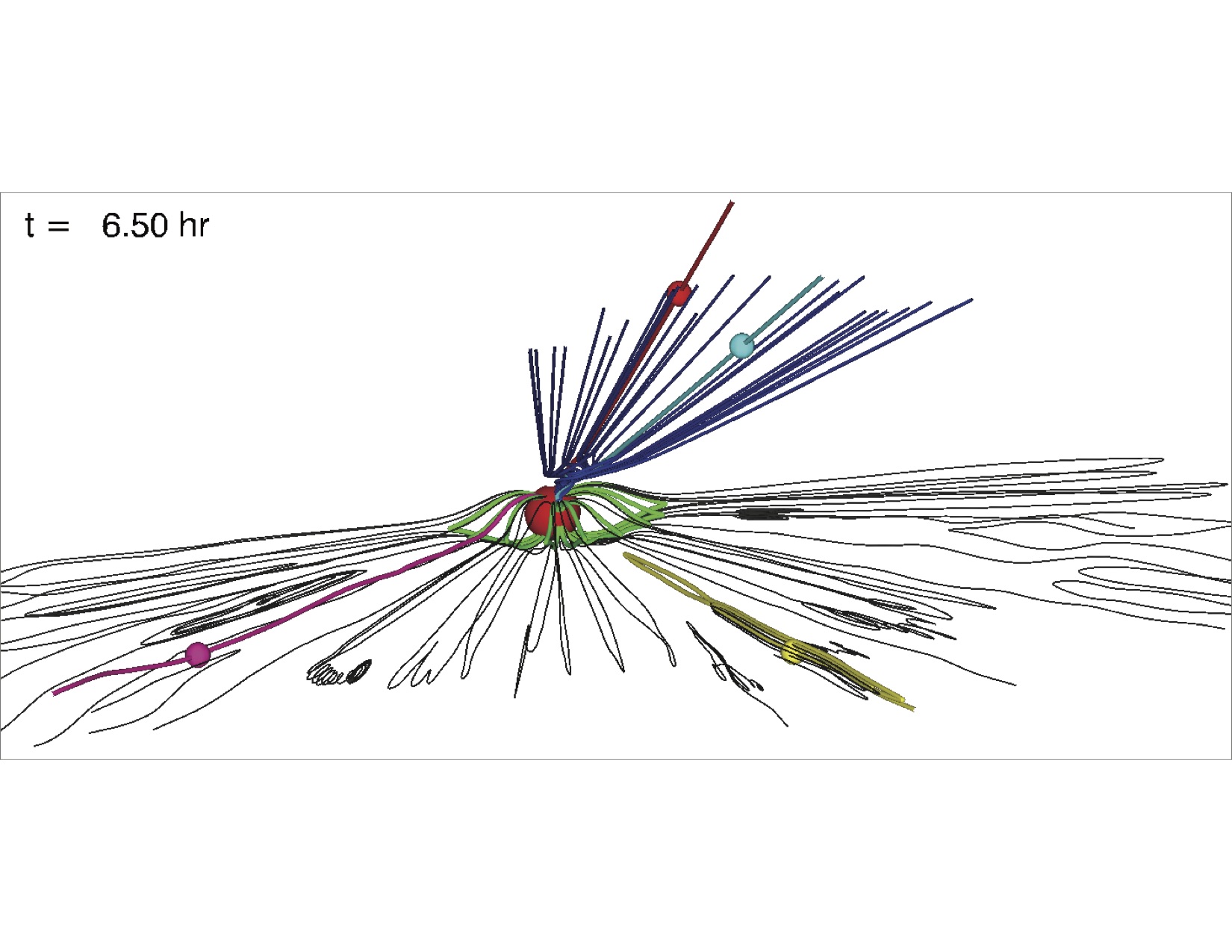}
	\includegraphics[width=0.5\textwidth]{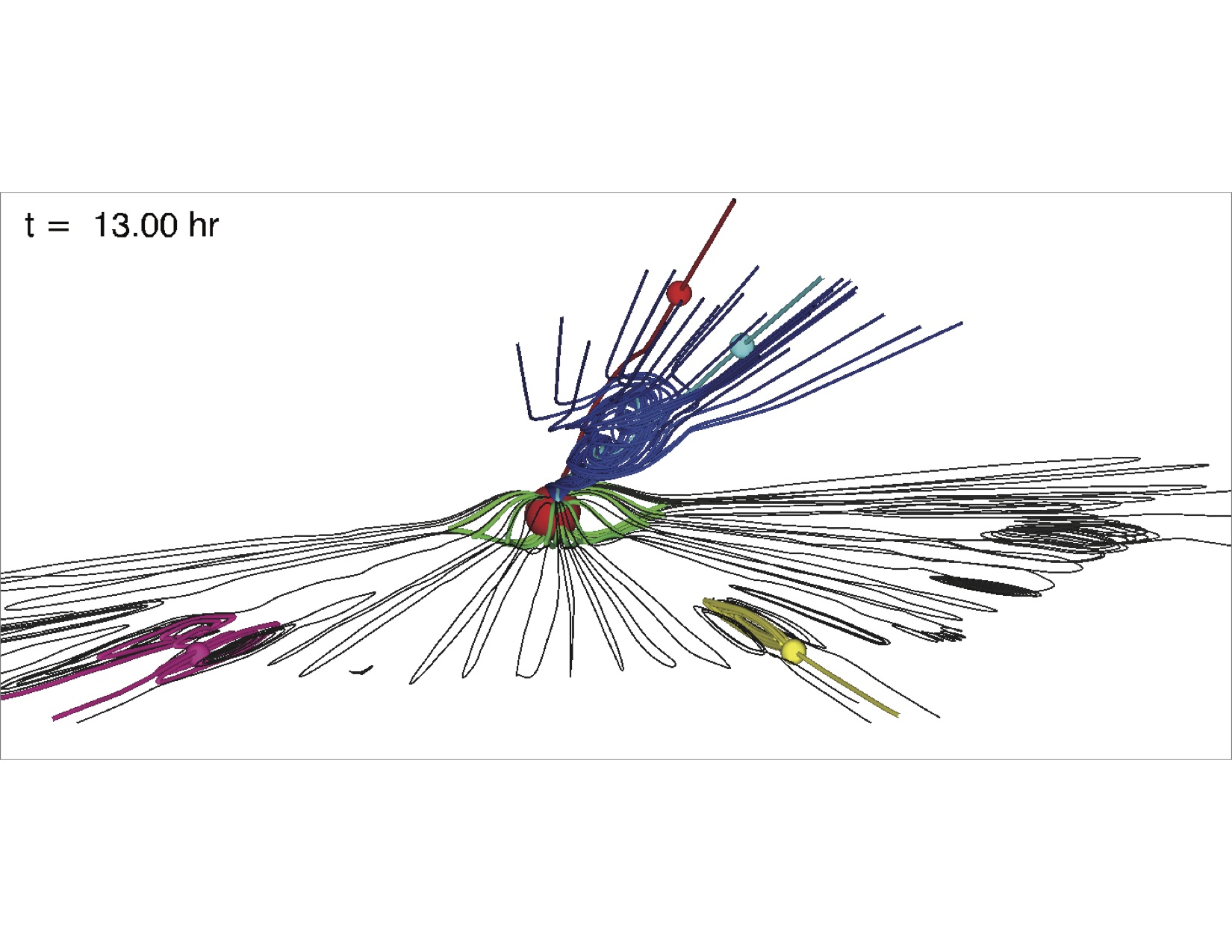}
	\includegraphics[width=0.5\textwidth]{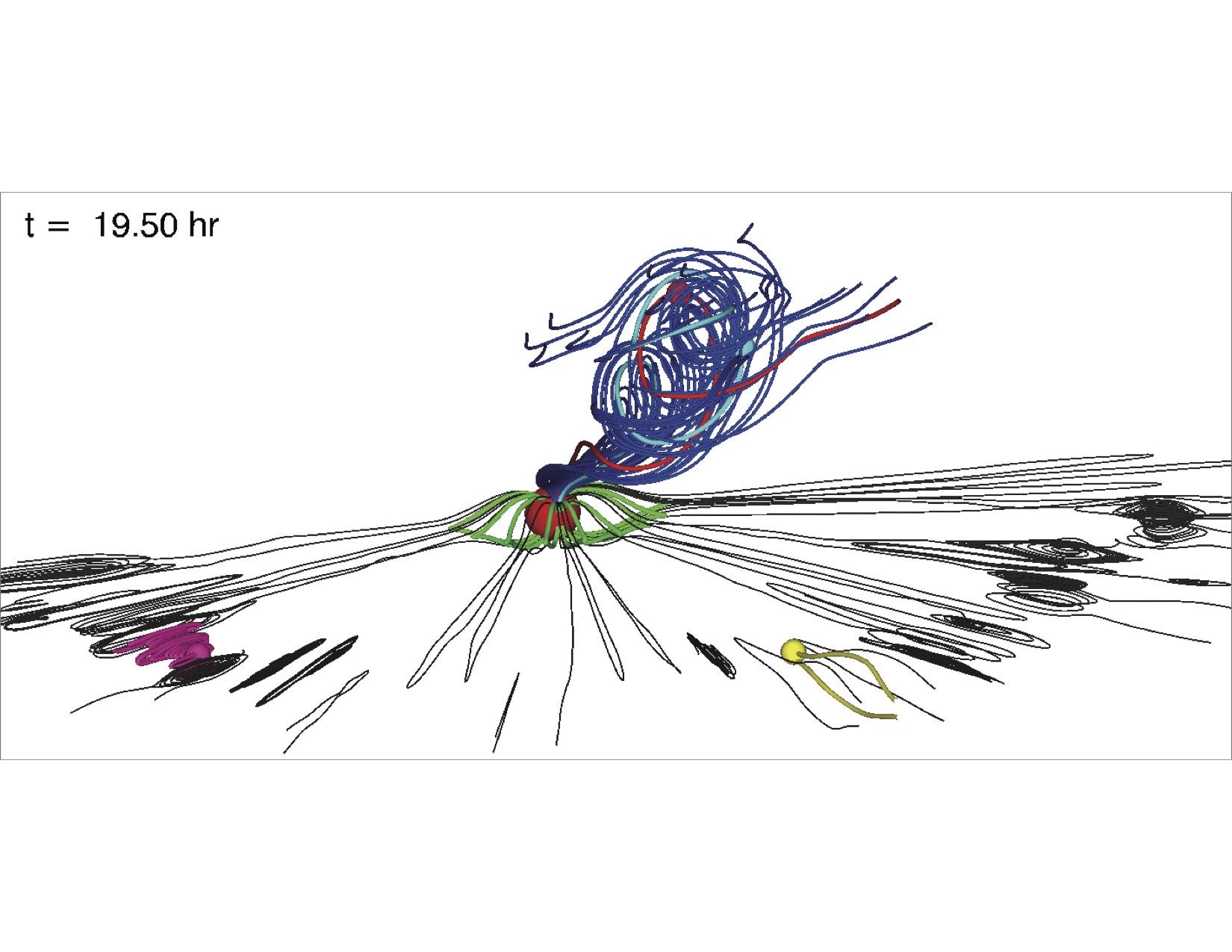}
	\includegraphics[width=0.5\textwidth]{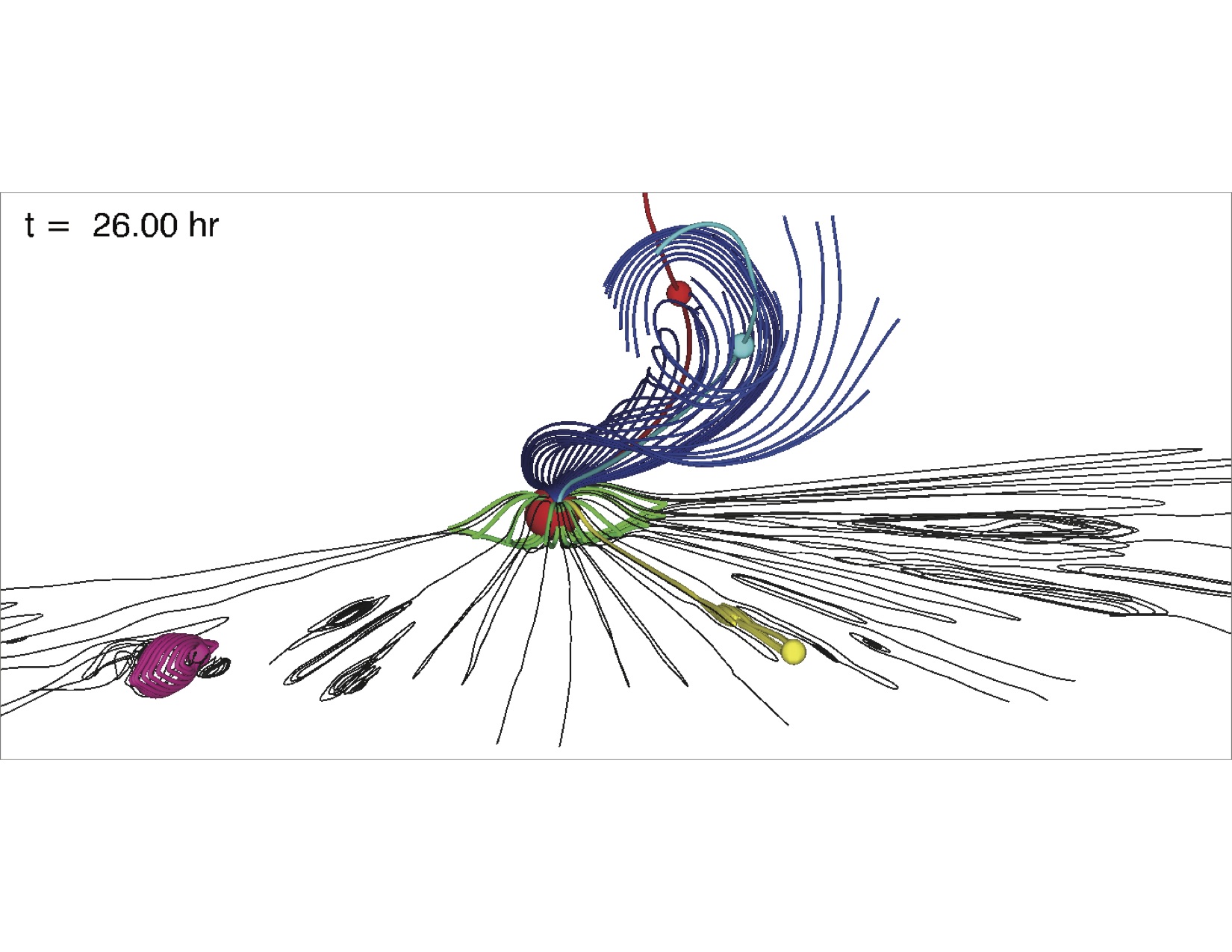}
	\includegraphics[width=0.5\textwidth]{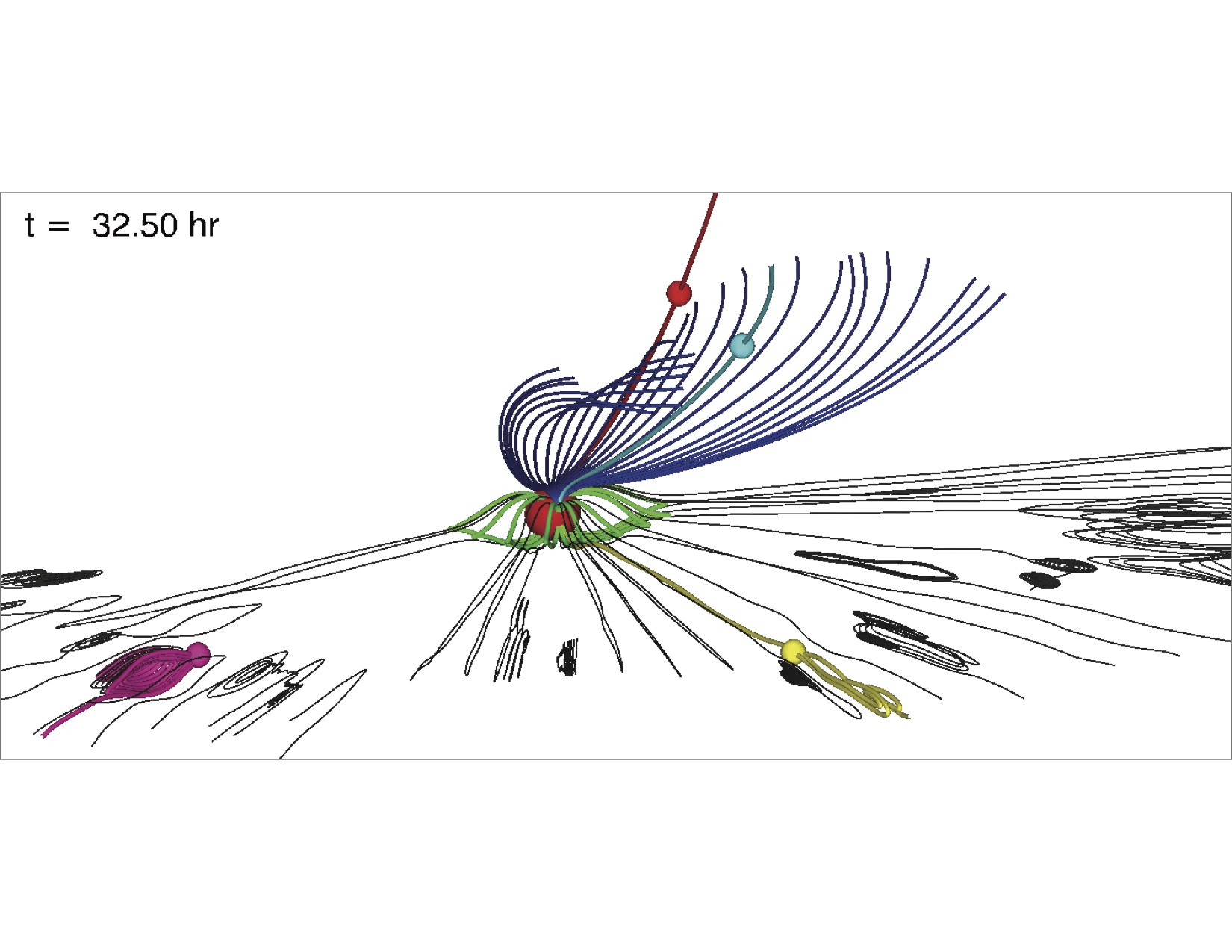}
	\caption{Six panels showing the evolution of streamer blobs
	in the HCS and the TAW. The black magnetic field lines show
	the interconnected blobs emanating from the top of the
	helmet streamer (shown in green). The blue field lines drawn
	from the surface show the S-Web arc and the TAW generated
	by the rotational photospheric driver. Four stationary
	observers and their magnetic field line connectivities are plotted in magenta,
	yellow, red, and light blue. \\ (An animation of this figure
	is available.)}
	\label{fig1}
\end{figure*}

This flow ($v_{max} < 9 km s^{-1}$) was applied to the lower boundary and rotated the field in a
100~Mm region by 180 degrees, effectively moving flux from one side
of the corridor to the other. The cyan dots drawn across Figure \ref{fig0}C represent the foot points of the 
blue magnetic field lines shown in Figure \ref{fig1}. The boundary motion of the magnetic
field introduces a significant twist component which
then propagates into the heliosphere along these open field lines as
a large-scale TAW.

\section{MHD Simulation Results}
\label{sxn_sim}


\subsection{System Evolution and the Magnetic Connectivity of Stationary Observers}
\label{ssxn_system}

The six frames of Figure~\ref{fig1} show the overall domain and
system evolution throughout our simulation (also discussed in
\citetalias{Higginson2017b}). The green magnetic field lines outline
the location of the helmet streamer, whose outer limit marks the
start of the heliospheric plasma sheet. The black magnetic field
lines, drawn from within the equatorial plane beyond 20 $R_\odot$,
show the stream of blobs emanating from the streamer top at all
longitudes continuously emitted throughout the simulation. This the first type of structured 
slow solar wind variability which is discussed in Section \ref{ssxn_mhdrxn}. When the
the top of the helmet streamer is stretched outwards by the solar
wind, it eventually becomes thin enough to reconnect, allowing the
top of the helmet streamer to pinch off and form a small flux rope,
which then moves outwards with the solar wind. The scale of these
blobs in our simulation is set by the scale of the magnetic
reconnection, which occurs at the scale of the grid. While the size
of these blobs will vary with Lundquist number in our simulation,
we will show that they have similar signatures to the streamer blobs
observed by \citet{Sheeley2009}, \citet{Rouillard2011}, and others.

Also shown in Figure~\ref{fig1}, rooted in the northern hemisphere
at the lower boundary, are blue magnetic field lines drawn across
the coronal-hole corridor. These field lines map out the high-latitude
S-Web arc before the driving takes place ($t = 0.00~hr$) and their
foot points do not move throughout the simulation. \citetalias{Higginson2017b} showed why 
this S-Web arc is predicted to also correspond to locations of slow solar wind, along with the HCS. When the rotational
motion takes place at the lower boundary between times $t = 0.00~hr$
and $t = 9.5~hr$, it launches an \Alfven wave onto the open magnetic
field lines. The start of this wave is visible at time $t = 6.50~hr$
and the subsequent snapshots in Figure~\ref{fig1} show its propagation
through the domain. By time $t = 32.50~hr$ most of this \Alfven
wave structure has left the domain through the outer boundary at
30 $R_\odot$. Note that while this wave slightly distorts the HCS,
the blobs shown within the heliospheric plasma sheet in Figure~\ref{fig1}
remain largely unaffected by the wave. We will argue that the
properties of our simulation's large-scale TAW are qualitatively
similar to the in situ observations of ``pseudo-flux rope'' waves
and that these structures can result, on a smaller scale than shown here, from interchange reconnection predicted to take place in regions which correspond to an S-Web arc \citepalias[e.g.][and references therein]{Higginson2017b},. This is the second type of structured slow
solar wind variability which is discussed in Section \ref{ssxn_mhdwave}.

Finally, Figure~\ref{fig1} also shows the location and magnetic
field line connectivity of four stationary observers marked by colored spheres.
The instantaneous magnetic field line from each of these points is
plotted in the same color scheme. The red (S1) and light blue (S2)
observers along the S-Web arc are located at
$\boldsymbol{r}_{S1}=(21R_\odot, 54^\circ, 0^\circ)$ and
$\boldsymbol{r}_{S2}=(21R_\odot, 60^\circ, 6^\circ)$ respectively.
S1 and S2 show the connectivity of a point in space as the TAW
passes over it, in contrast to the dark blue field lines which show
the magnetic field connectivity drawn from the surface. The magenta
(E1) and yellow (E2) observers lie in the ecliptic plane, located
at $\boldsymbol{r}_{E1}=(21R_\odot, 90^\circ, -53^\circ)$ and
$\boldsymbol{r}_{E2}=(21R_\odot, 90^\circ, 7^\circ)$ respectively.
They sample the 3D flux ropes that form in the HCS which correspond
to the well-known streamer blobs. We encourage the reader to see
the animation of Figure~\ref{fig1} that is included as an online
electronic supplement to the article.


\subsection{Variability in the HCS-Associated Wind: Generation and Propagation of Small-Scale Flux Ropes}
\label{ssxn_mhdrxn}


\begin{figure*}
	\includegraphics[width=1.0\textwidth]{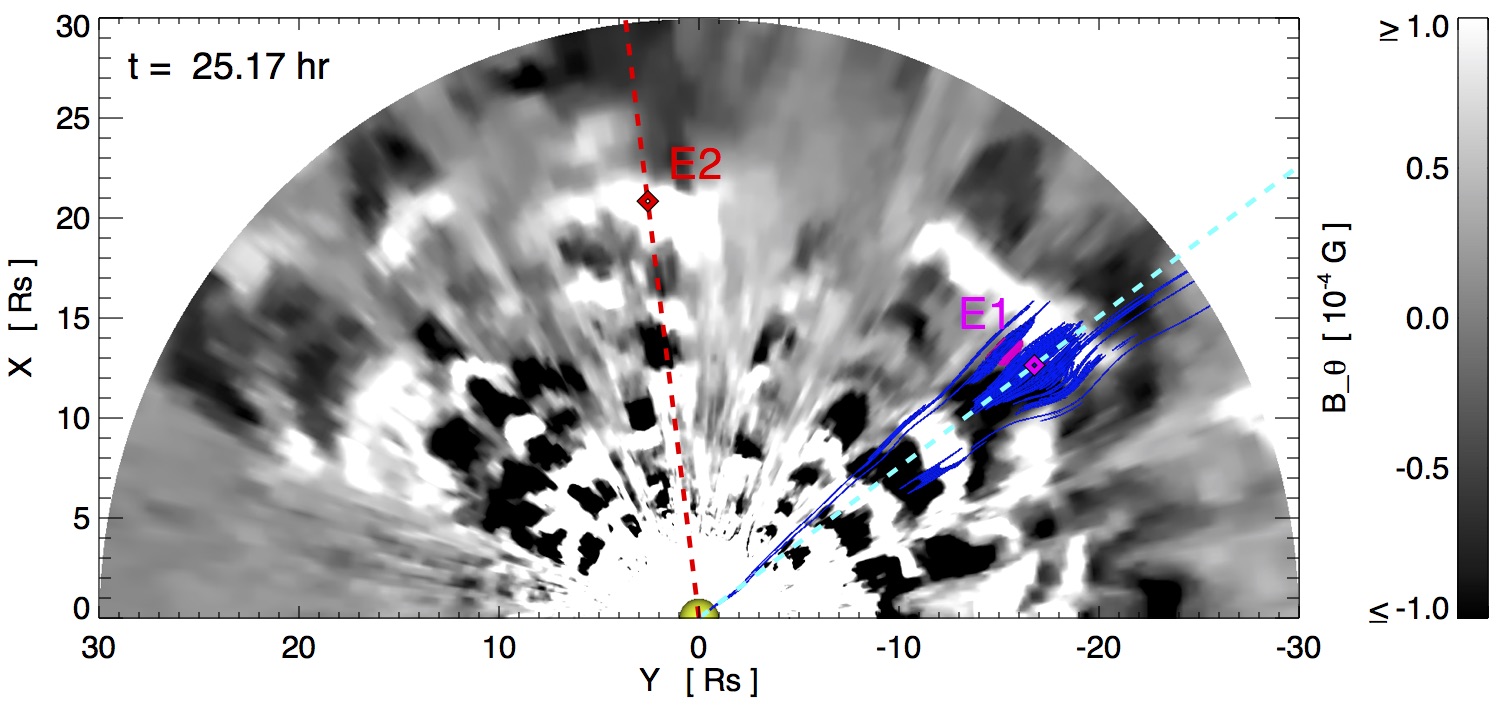}
	\caption{An equatorial cut of $B_\theta$ viewed from above.
	The greyscale is heavily saturated such that it shows the
	sign of $B_\theta$. Each pair of black and white signals
	show the location of a blob in the HCS. Also shown are
	observers E1 and E2 and their radial vectors. Observer E1
	is accompanied by blue magnetic field lines drawn for
	context. \\ (An animation of this figure is available.)\\ }
	\label{fig2}
\end{figure*}

Figure~\ref{fig1} {illustrates} the dynamic and 3D nature of the blobs
originating from the top of the helmet streamer. 
{Figure~\ref{fig2}, an equatorial cut of $B_\theta$ where the viewer is looking
down on the HCS from the north pole, shows how these blobs fill the heliospheric plasma sheet.}
By showing $B_\theta$ in
saturated greyscale, we highlight the locations where $B_\theta$
is either into or out of the page (i.e. northward or southward
toward either pole). The leading edge of the structures have a $B_\theta$ component in
one direction, while the trailing edges have the opposite, as expected from small flux ropes. In
Figure~\ref{fig2}, the leading edges have a positive $B_\theta$
(towards the north pole) and are shown in white. The blobs show no
longitudinal preferences and move outwards in a continuous radial
stream. Also shown in Figure~\ref{fig2} are the locations of two
of the stationary observers E1 (magenta) and E2 (now shown in red)
from Figure~\ref{fig1}. The dotted lines shown in light blue and
red represent the rays drawn from the Sun through the observers
which were used to generate the simulated remote observations below.
Blue magnetic field lines are also shown around E1 for context.

Clearly visible in Figure~\ref{fig2} is the longitudinal extent of
the blobs. While the flux rope structures observed by \citet{Sheeley2009}
can be tens of degrees wide, here we see smaller longitudinal widths
of less than ten degrees. This is due to the negligible longitudinal guide field
present in our symmetric helmet streamer. As the streamer top
stretches out and eventually pinches off, forming a small flux rope,
the shear component at the top of the helmet streamer becomes the
core field of the flux rope, oriented perpendicular to the radial
direction. In our simulation this component is generally small.
Small longitudinal widths are consistent, however, with the
multi-spacecraft in situ observations of small flux ropes in the
solar wind by \citet{Kilpua2009b}. They observed that spacecraft
separated by only a few degrees would often not encounter the same
transient structure, and even if they did, the features could be
quite different, suggesting a small width and high level of complexity
within the flux ropes.


\begin{figure*}
	\includegraphics[width=1.0\textwidth]{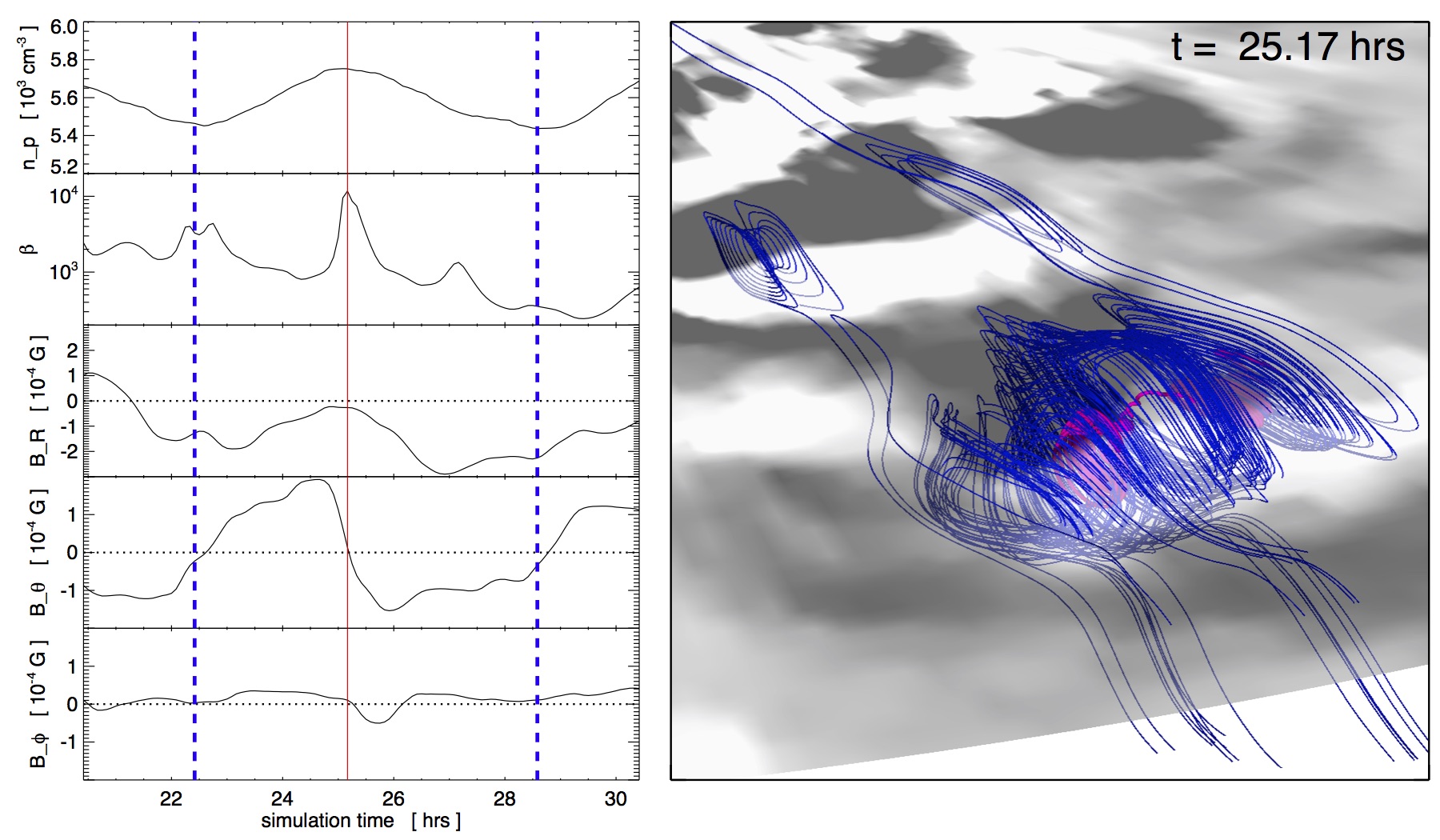}
	\caption{Left panel: in situ data detected by observer E1
	corresponding to a single, well-structured streamer blob
	flux rope event (denoted by the vertical dashed lines).
	Right panel: the zoomed-in, 3D view of E1 from Figure~\ref{fig2}
	showing a portion of the magenta magnetic field line traced
	from E1 and the surrounding blue field lines showing the
	HCS flux rope structure.  \\ (An animation of this figure
	is available)\\ }
	\label{fig3}
\end{figure*}

In Figure~\ref{fig3} we take a closer look at the structure of a
single streamer-blob flux rope. The right panel shows a frame from
the animation of Figure~\ref{fig2}, but now the viewer is looking
edge on to the equatorial plane and watching the flux rope as it
moves towards the outer boundary in the bottom right corner. The
black and white transparent surface is the same $B_\theta$ from
Figure~\ref{fig2} as are the magenta and blue magnetic field lines
surrounding the observer E1. Here we also see, especially in the
animation of this figure, that these flux ropes are not isolated,
disconnected structures, but rather that the magnetic connectivity
often threads multiple flux ropes. This is due to the three dimensionality of the
 pinching-off reconnection of the helmet streamer, in particular,
  different longitudes pinch-off reconnect at different
times, meaning they retain some partial magnetic connection to
nearby regions. Because each event is fully 3D and different than
its neighbors, the overall effect is that of interconnectedness,
rather than fully independent and separate flux rope structures.
This could also be due to, or exacerbated by, dynamic magnetic
reconnection which takes place between solar wind structures as the
blob moves outwards in the solar wind, as discussed by
\citet{Gosling2006hcsrxn} in observations and \citet{Higginson2017a}
in simulations.

In the left panel of Figure~\ref{fig3} we plot the in situ parameters
observed by E1 as the flux rope sweeps across it. From top to bottom
we plot number density ($n_p$), plasma beta ($\beta=8\pi P/B^2$),
and the three components of the magnetic field ($B_r$, $B_\theta$,
$B_\phi$), where $B_r$ is towards or away from the Sun, $B_\theta$
is towards the north or south pole, and  $B_\phi$ completes the
right-handed coordinate system. The red line indicates the time
shown in the right panel, and the blue dashed lines bound the
event. Our flux rope blobs move outwards with the solar wind and
show a smooth and organized rotation in the magnetic field. Here,
$B_\theta$ starts out positive before flipping smoothly through
zero at the core of the structure to become negative on the other
side. Our density shows a slight increase, on the order of 6\%, at
the center of the flux rope. The magnitude of this density fluctuation
is almost certainly underestimated with our isothermal solar wind model.
Additionally, the behavior of the plasma $\beta$ in our simulation
flux rope is opposite that reported in the literature, due to our
lack of a flux rope core (axial) field. Our magnetic field effectively
goes to zero at the center of the structure, causing the spike in
plasma $\beta$. However, the bipolar magnetic field rotation signature
in our streamer-blob flux ropes closely resembles those reported
by \citet{Cartwright2008}.


\begin{figure*}
	\centerline{ \includegraphics[width=0.64\textwidth]{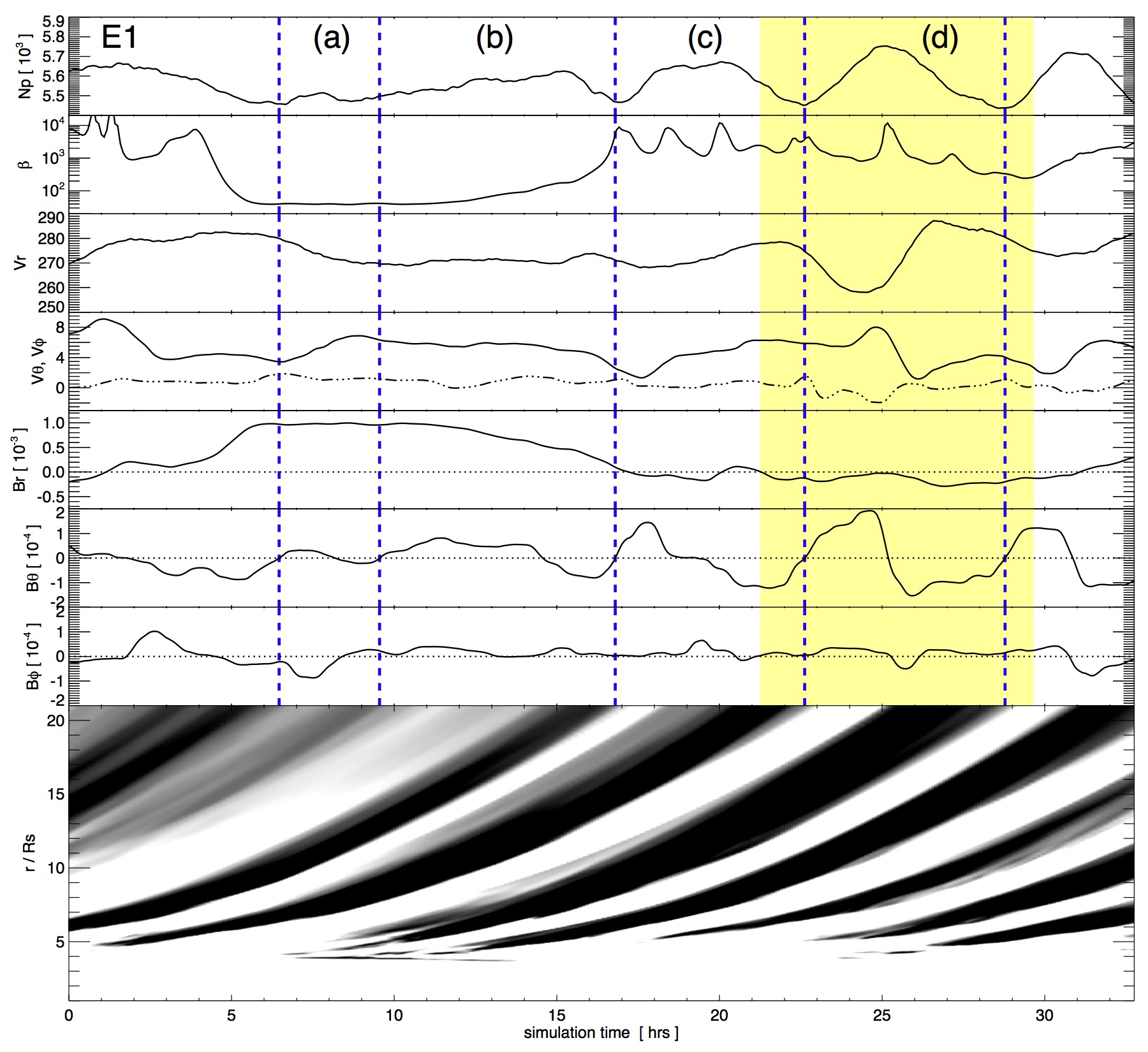} }
	\caption{The full time series of $N_p$, $\beta$, $\boldsymbol{V}$,
	$\boldsymbol{B}$ at E1. The bottom panel shows the height-time
	J-map \citep[after][]{Sheeley1999} of the Figure~\ref{fig2}
	${B_\theta}$ quantity that define the 3D flux rope boundaries
	(also shown as dashed vertical blue lines).\\ }
	\label{fig4}
\end{figure*}


\begin{figure*}
	\centerline{ \includegraphics[width=0.64\textwidth]{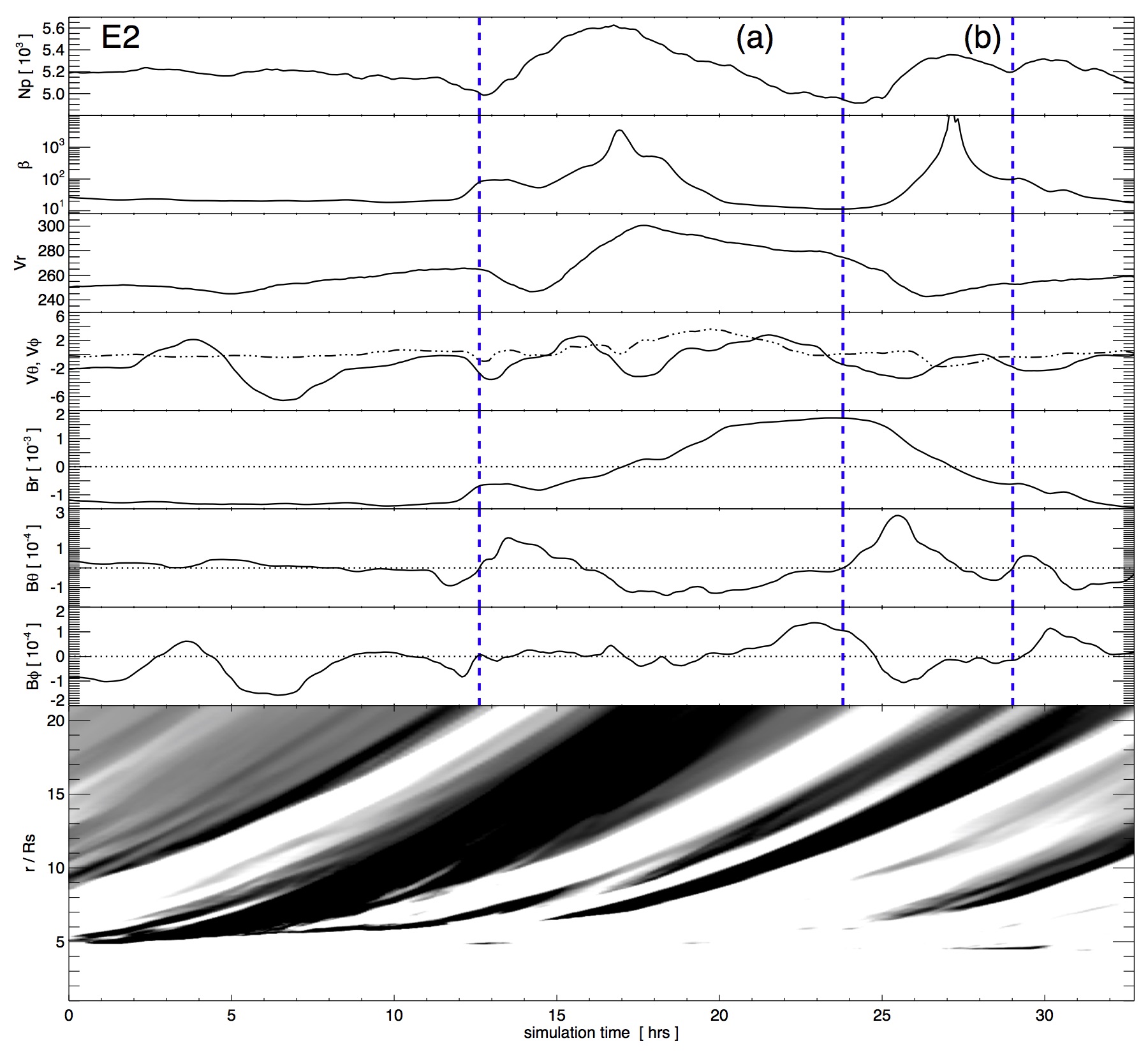} }
	\caption{The E2 plasma and magnetic field time series and
	J-map, in the same format as Figure~\ref{fig4}.\\}
	\label{fig5}
\end{figure*}

Figure~\ref{fig4} shows the in-situ parameters number density
($N_p$), plasma beta ($\beta$), radial velocity ($V_r$), latitudinal
and longitudinal velocity ($V_\theta$ and $V_\phi$), and the three
components of the magnetic field ($B_r$, $B_\theta$, $B_\phi$)
detected at E1 for $\sim$33 hours of the simulation. The bottom
panel shows a height-time J-map \citep[after][]{Sheeley1999} of
$B_\theta$ along the dotted light blue line from Figure~\ref{fig2}
at the position angle of E1, where the vertical axis is height above
the solar surface and simulation time is along the horizontal axis.
The time period shown in Figure~\ref{fig3} is highlighted in yellow
in Figure~\ref{fig4}. The top of the J-map is set at the radial
height of E1 (21 $R_\odot$) so that the top of the J-map aligns
with the in situ data plotted above it. The blue dotted lines from
Figure~\ref{fig3} which bound the event are also plotted in
Figure~\ref{fig4}. Additional blue lines are plotted to show the
boundary between flux ropes, defined as the transition from negative
to positive $B_\theta$. Figure~\ref{fig4} cleanly illustrates the
steady and continuous nature of the blobs, their smooth motion
outwards with the solar wind velocity (seen in the J-map), and also
shows how magnetically complex the structures can be (in situ data)
while retaining their smooth rotations in magnetic field.

Figure~\ref{fig5} shows the same quantities as shown in Figure~\ref{fig4}
for the observer E2. The radial cut used for the J-map in the bottom
panel is indicated by the red dotted line in Figure~\ref{fig2}.
Comparing Figures~\ref{fig2}, \ref{fig4}, and \ref{fig5} it is clear
that the blobs occur at all longitudes and have similar properties.
The radial extent of some of the blobs in Figure~\ref{fig5} varies
more than the blobs in Figure~\ref{fig4} because in this location
the HCS becomes slightly distorted, developing a slight ``ballerina
skirt ruffle,'' and is shifted to slightly lower latitudes. Rather
than cutting straight through the center of the blobs, the E2
observer essentially skims across the top of the HCS structures
instead. This also seen in the $B_r$ time series---$B_r$ is a constant, 
negative polarity, until the first flux rope boundary at $t\sim12.5$~$hr$ 
and afterwards, $B_r$ shows two consecutive HCS crossings
and three easily-identified streamer-blob  flux ropes. The movement
of the HCS is due to the driving and resulting dynamics in the
simulation, which will be discussed in Section~\ref{ssxn_mhdwave}
below.


\subsection{Variability in the S-Web-Associated Wind: Generation and Propagation of a Large-Scale Torsional Alfv\'{e}n Wave}
\label{ssxn_mhdwave}

As described in \citetalias{Higginson2017b}, to simulate the effect
of supergranular driving on an S-Web coronal-hole corridor we placed
a $1 \pi$ rotation in $(\theta,\phi)$ on the solar surface, which
overlapped with the closed magnetic field on both sides of the
coronal-hole corridor. This motion displaced the coronal hole
boundary and generated a plethora of interchange reconnection as
the boundary relaxed to its new equilibrium state. This
interchange reconnection is responsible for the release of coronal
plasma from the closed field regions onto open field lines all along
the S-Web arc \citepalias{Higginson2017b} and is the origin of the
slow solar wind found far the heliospheric current sheet. The effect
of this photospheric rotation on the open magnetic field was to
launch a large-scale \Alfven wave which propagated outwards into
the heliosphere.


\begin{figure*}
	\centerline{ \includegraphics[width=1.0\textwidth]{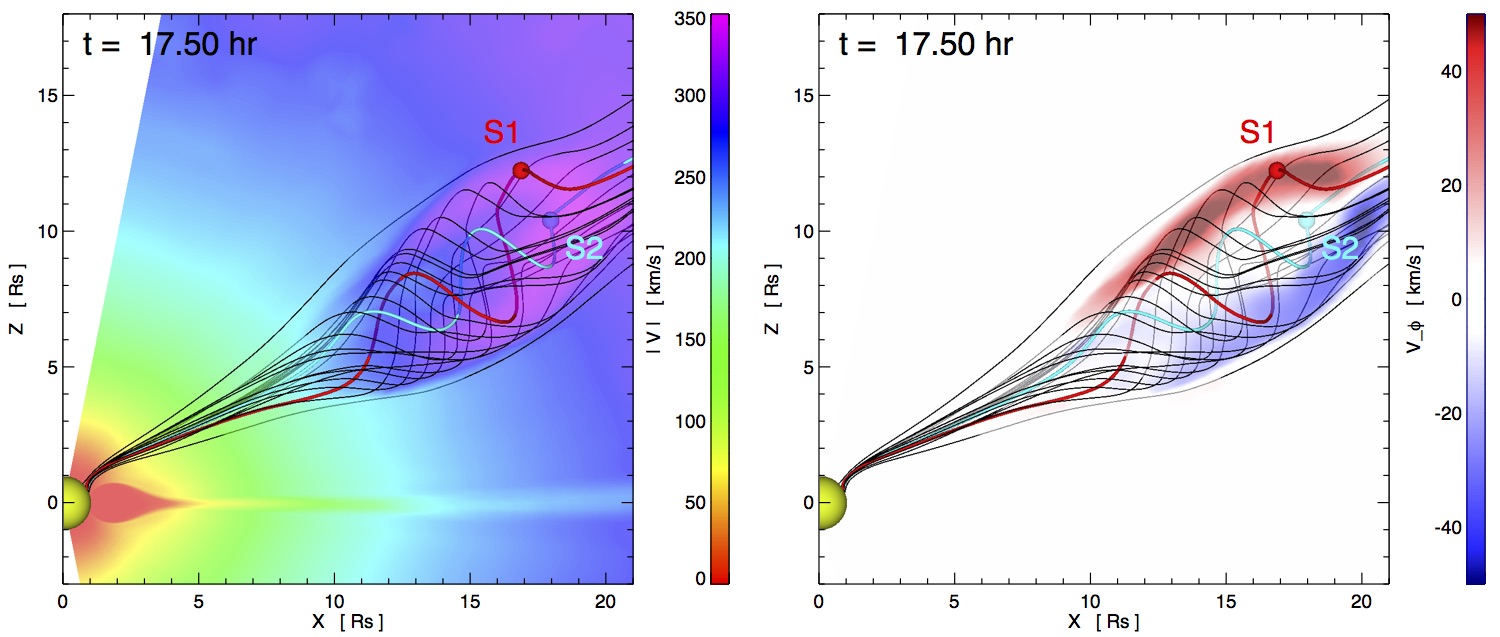} }
	\caption{Longitudinal cut at $\phi = 0$ of $|V|$ (left) and
	$V_\phi$ (right) showing the TAW propagating through the S1 (red), S2 (light
	blue) observation points. Representative surrounding magnetic field lines are 
	also shown in black to illustrate the ``pseudo-flux rope'' structure of the TAW.
	\\ (An animation of this figure is available) \\}
	\label{fig6}
\end{figure*}

Figure~\ref{fig6} shows this TAW after the surface rotation has
stopped and the wave has propagated out to $\sim$15 $R_\odot$. The
left panel shows contours of velocity magnitude ($|V|$) in a slice
through zero longitude. The right panel shows contours of $V_\phi$
(into and out of the page) at the same locations. The black magnetic
field lines are shown for context around the two observers S1 and
S2, originally shown in Figure~\ref{fig1}.  The animation included
online shows the generation and propagation of this wave for the
whole simulation period. Here in this 3D view, the resemblance of
this structure to a flux rope is on display; the magnetic field
lines seem to coil around an axial magnetic field in the direction
of propagation.


\begin{figure*}
	\centerline{ \includegraphics[width=0.64\textwidth]{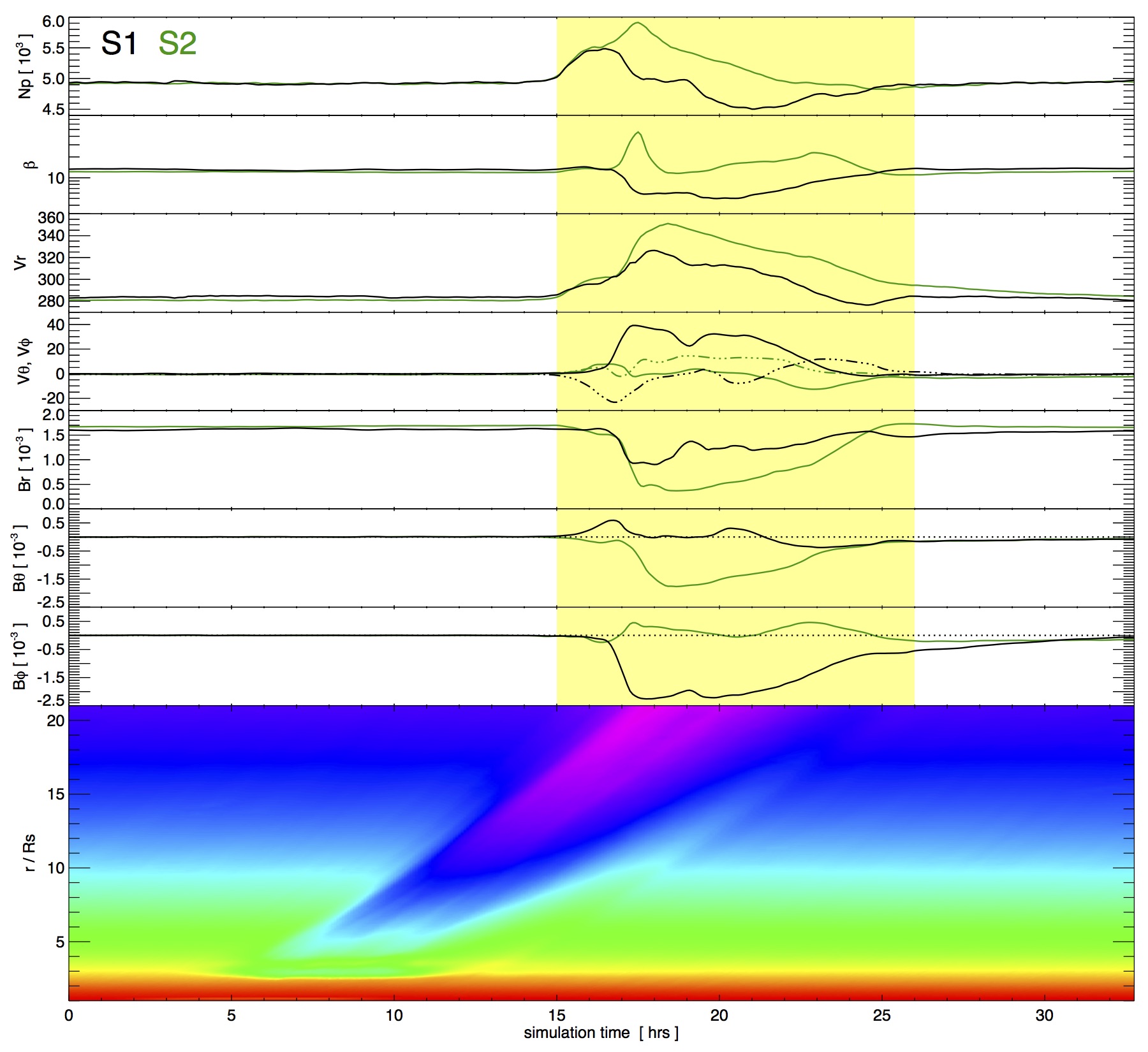} }
	\caption{The full time series of $N_p$, $\beta$, $\boldsymbol{V}$,
	$\boldsymbol{B}$ at S1 (black) and S2 (green) observation points. 
	The height-time J-map plot through S1 of ${|V|}$ shows the radial 
	propagation and evolution
	of the large-scale TAW.\\}
	\label{fig7}
\end{figure*}

To compare this structure to the flux ropes seen in the HCS, we
again plot the in situ data from the observers in Figure~\ref{fig7}.
Here we show the time series for the full $\sim$33 simulated hours of
$N_p$, $\beta$, $\boldsymbol{V}$, $\boldsymbol{B}$ at S1 (black) and
S2 (green) in panels 1-7. 
%
%
The bottom panel
shows the J-map of $|V|$ using a ray through S1 and the color scale
from Figure~\ref{fig6}. As before, the top of the J-map lines up
with the location of the observer for comparison to the in situ
time series. The time period during which the TAW is passing over
S1 is highlighted in yellow. Immediately obvious is the scale of
the transient above the background solar wind. The density and all
three of the velocity components show much larger enhancements
than the flux rope structures. This is due in part to the unrealistically
large driving motion which we used at the surface due to numerical
constraints, however, we will argue below that this structures of this type should
be visible at smaller scales in the solar wind. Examining the
magnetic field components we see that while the flux rope structures
exhibited a smooth rotation through zero, this structure shows a
rotation offset from zero in $B_\phi$ for S1 and $B_\theta$ for S2.
The S1 and S2 observation points are at the center and edge of the wave structure in 
longitude respectively but are roughly equivalent distances from the central radial axis of 
the large-scale propagating TAW disturbance. Hence, the S1, S2 profiles of $B_\phi$ and $B_\theta$ 
represent an $\sim$90$^\circ$ phase-shift in the fluctuation quantities.
While multipoint in situ measurements would certainly help to differentiate 
between TAWs and small-scale flux ropes, the magnetic field rotation offset 
from zero signature by itself is generally not sufficient. 
This is even more relevant since we know in our simulation that the observers 
are traversing the TAW structure axially instead of cutting across its diameter 
as in the case of our flux ropes in the HCS.


\begin{figure*}
	\centerline{ \includegraphics[width=0.90\textwidth, height=0.27\textwidth]{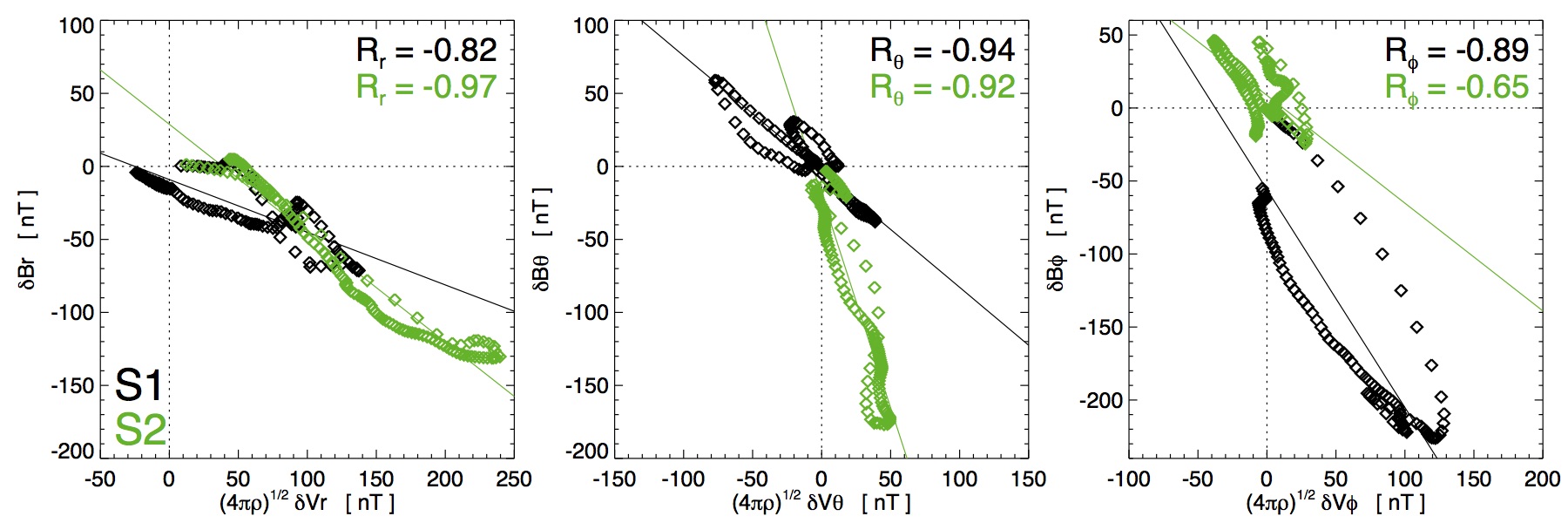} }
	\centerline{ \includegraphics[width=0.90\textwidth, height=0.27\textwidth]{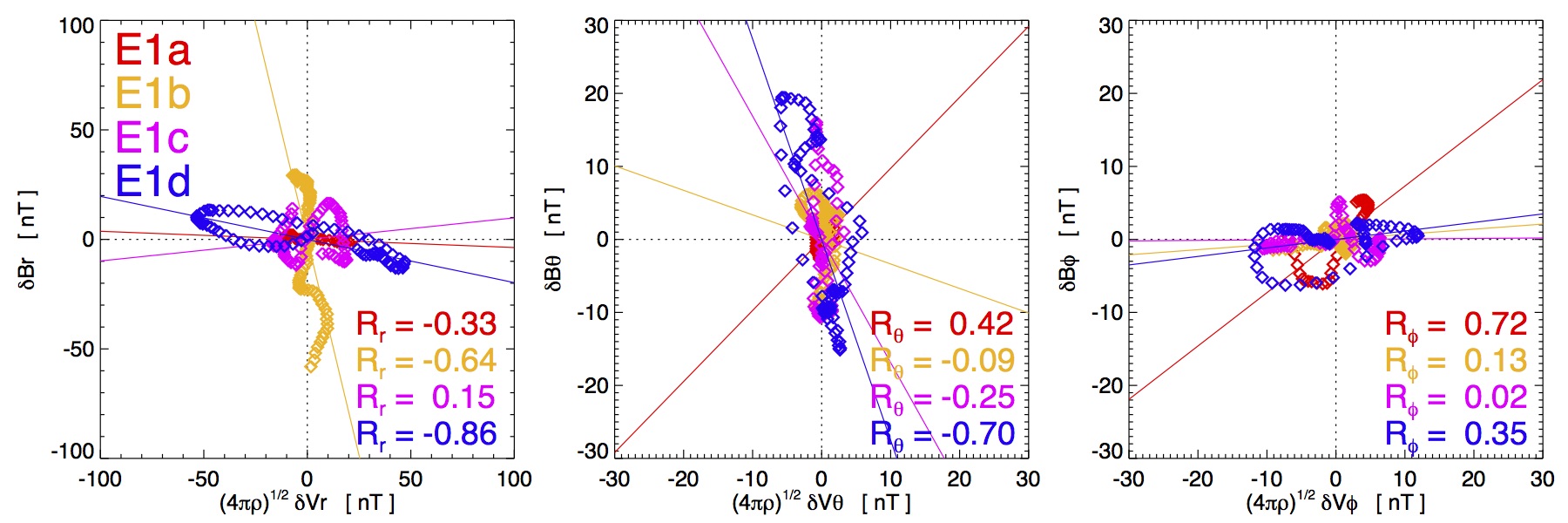} }
	\centerline{ \includegraphics[width=0.90\textwidth, height=0.27\textwidth]{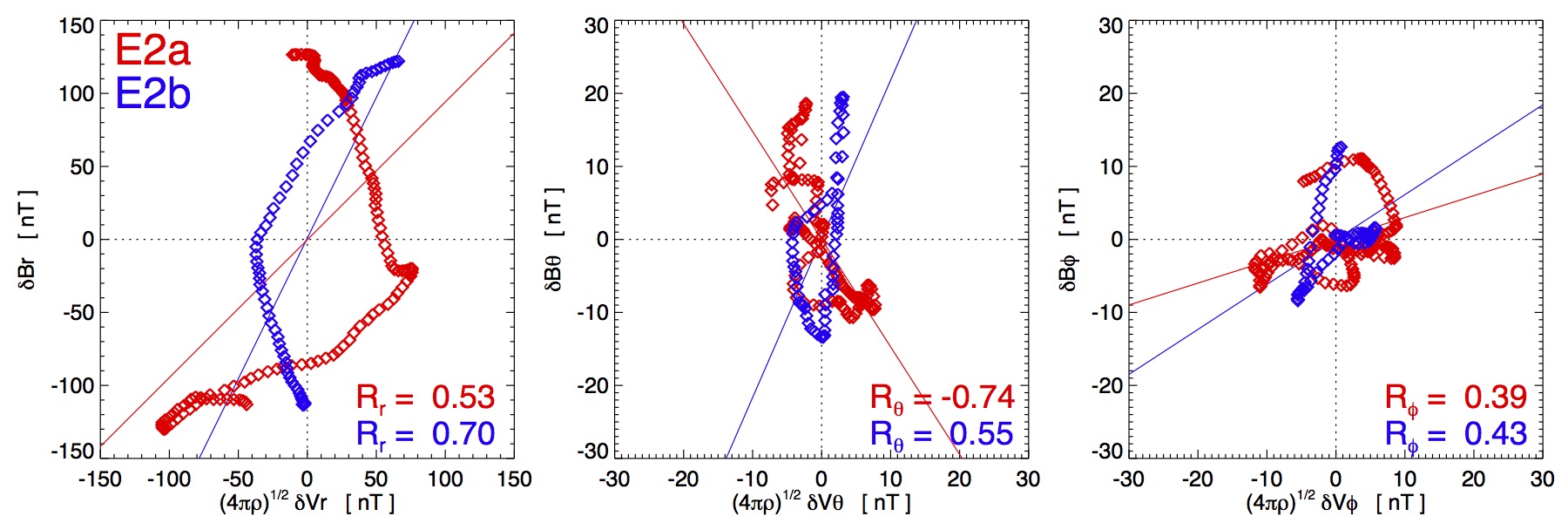} }
	\caption{Left, middle, and right panels show the Alfv\'{e}nicity
	plot of $\delta \boldsymbol{B}$ vs. $(4\pi\rho)^{1/2} \; \delta
	\boldsymbol{V}$ in their $r$, $\theta$, and $\phi$ components,
	respectively for each of the events. The linear fits are plotted with solid colored lines and the correlation coefficients are listed. The TAW event observed by S1 (black) and S2 (green) are shown in the top row. The four flux-rope blobs observed by observer E1 are shown in red, yellow, pink, and blue in the middle row. The two flux-rope blobs observed by E2 are shown in the bottom row in red and blue. The high correlation coefficients of TAW event in the top row show that it is highly Alfv\'{e}nic. } 
	\label{fig8}
\end{figure*}

To differentiate between a flux rope and a TAW we must examine the
\Alfvenicity of the structure during the time period highlighted in yellow in Figure~\ref{fig7}. The TAW identification criteria for in situ data used by \citet{Yu2014,Yu2016} is the magnitude of the correlation between the magnetic field fluctuations $\delta \boldsymbol{B}$ and the density-normalized velocity fluctuations $(4\pi\rho)^{1/2} \; \delta\boldsymbol{V}$. The fluctuations are calculated in the usual fashion as $\delta \boldsymbol{B} \equiv \boldsymbol{B}-\langle \boldsymbol{B} \rangle$ and $\delta \boldsymbol{V} \equiv \boldsymbol{V}-\langle \boldsymbol{V} \rangle$ where the mean value is the average (or a sufficiently long running-average) of the time series. The period of interest is considered Alfv\'{e}nic if all three of the fluctuation vector components have correlation magnitudes of $|R| > 0.5$ or if two components are strongly correlated with $|R| >0.6$ and the third component is at least weakly correlated with $|R| > 0.3$.  

Figure~\ref{fig8} plots $\delta \boldsymbol{B}$ against $(4\pi\rho)^{1/2} \; \delta
\boldsymbol{V}$ for each TAW and flux rope event. The fluctuation vectors are separated into their respective $r$-components (left panel), $\theta$-components (middle panel) and $\phi$-components (right panel). The TAW observed by S1 (black) and S2 (green) is analyzed in the top row. The middle row shows the analysis of the four flux rope events observed by E1 (as delineated in Figure~\ref{fig4}) in red, yellow, pink, and blue, and the bottom row shows the two flux rope events observed by E2 (as delineated in Figure~\ref{fig5}) in red and blue. The correlation coefficients for each component in each interval are listed in each plot. The linear fits to the component pairs are also shown in each panel as dashed lines. For observer S1 (S2) which measured the TAW, we obtained correlation magnitude values that far exceed the observational Alfv\'{e}nic threshold criteria, as expected: {$|R_r|$ = 0.82 (0.97), $|R_\theta|$ = 0.94 (0.92), and $|R_\phi|$ = 0.89 (0.65)}.  Each of the HCS flux rope {event intervals (E1a$-$d, E2a,b) fail to pass the empirical in-situ} criteria for being considered Alfv\'{e}nic.

We also computed the Wal\'{e}n number {$R_W(t)$} \citep{walen44} throughout each event interval, defined as
\begin{equation}
\big \langle R_W \big \rangle \equiv \bigg \langle \frac{(4\pi \rho)^{1/2} \delta V_{\perp}}{\delta B_{\perp}} \bigg \rangle ,
\end{equation}
where $\delta V_{\perp} = (\delta V_\theta^2 + \delta V_\phi^2)^{1/2}$, $\delta B_{\perp} = (\delta B_\theta^2 + \delta B_\phi^2)^{1/2}$, and the $\langle \cdot \rangle$ brackets denote the event-average of the synthetic in-situ time series. An ideal Alfv\'{e}n wave in an isotropic plasma would have $R_W = 1$. The minimum, maximum, {mean, and standard deviation} of $R_W$ {measured} by each observer over the course of each event are listed in Table~\ref{tab:dhdt}. Here we see that each {HCS flux-rope interval (event type ``blob'')} has a much higher, non-Alfv\'{e}nic maximum $R_W$ value than either of the {TAW event intervals (type ``wave'')}, and that the wave events show $R_W(t)$ distributions closer to {$\lesssim1$ than the blob events.}

\begin{figure*}
	\centerline{ \includegraphics[width=1.0\textwidth]{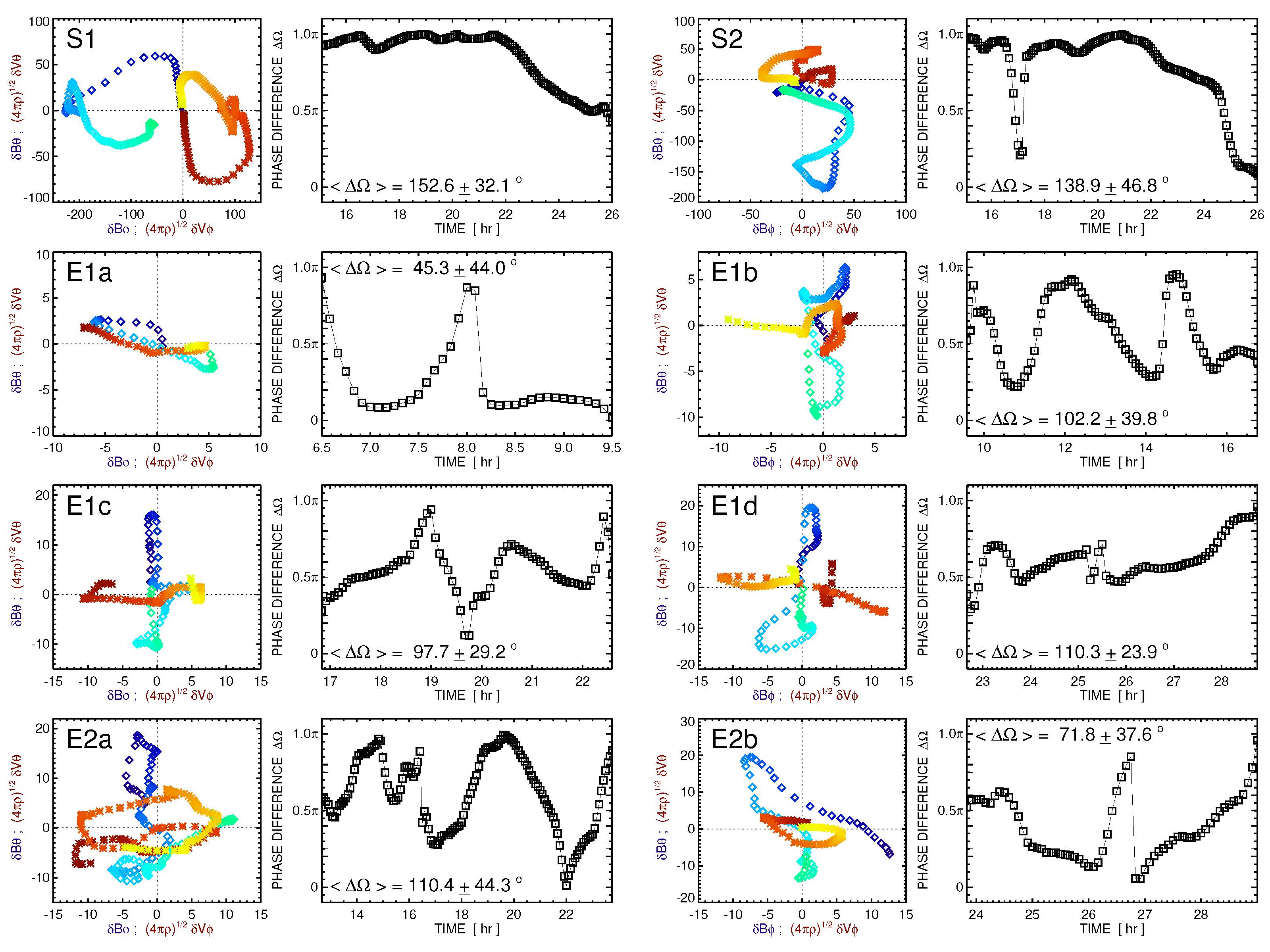} }
	\caption{Hodograms and phase angle evolution of the perpendicular components of the magnetic field $\delta \boldsymbol{B_\perp}$ and normalized velocity fluctuations $(4\pi\rho)^{1/2} \delta \boldsymbol{ V_\perp }$ in our TAW samples S1, S2 (top row), and each of the equatorial streamer blob events, E1a-d (second and third rows) and E2a,b (forth row). The colors in each hodogram show the temporal evolution of the magnetic fluctuations (blue-to-green) and the velocity fluctuations red-to-yellow over the duration of the events. The relative phase difference $\Delta \Omega$ between the $\delta \boldsymbol{ B_{\perp}}$ and $\delta \boldsymbol{ V_{\perp} }$ components are plotted next to each hodogram (also summarized in Table \ref{tab:dhdt})}.
	\label{fig9}
\end{figure*}

The analysis of the Wal\'{e}n numbe and the component correlation coefficients (Figure \ref{fig8}) analyze only the fluctuation magnitudes, so we also look at the the relative phase difference between the $\delta \boldsymbol{B_{\perp}}$ and $\delta \boldsymbol{V_{\perp}}$ components to provide additional information about the Alfv\'{e}nicity of our intervals. For a radial propagating disturbance, the fluctuation phase angles are defined as 
\begin{equation}
\alpha_B = \sin^{-1}\left[ {\frac{\delta B_\theta}{ \delta B_{\perp}}} \right] \; , \;\;\; \alpha_V = \sin^{-1}\left[ {\frac{\delta V_\theta}{ \delta V_{\perp}}} \right] \; ,
\end{equation}
which yield a relative phase difference of
\begin{equation}
\Delta \Omega = \left \{ \begin{array}{lcl}
  			\left | \alpha_B - \alpha_V \right | \;\; & \mathrm{for} \; & \left | \alpha_B-\alpha_V \right | \le \pi  \\
  			2\pi - \left | \alpha_B - \alpha_V \right | \;\; & \mathrm{for} \; &  \left | \alpha_B-\alpha_V \right | > \pi
			 \end{array} \right. ,
\end{equation}
with $\Delta \Omega \in [0,\pi]$ for $\alpha_B$, $\alpha_V$ $\in [ 0, 2\pi ]$. Here an ideal Alfv\'{e}n wave will have a phase difference between the field and velocity components of $\Delta \Omega = \pi = 180^{\circ}$. These quantities are also listed in Table \ref{tab:dhdt} and plotted in Figure~\ref{fig9}, along with the pictorial hodograms.

Figure \ref{fig9} shows the hodogram and relative phase angle evolution for each of the observed structures. The TAW observations by S1 and S2 are in the top row, the blob observations from E1a-d are in the second and third rows, and the blob observations from E2a,b are in the bottom row. The red-to-yellow (blue-to-green) transition in the hodograms on the left shows the time evolution in the velocity (magnetic) fluctuations, and the phase angle between these at each time is plotted in black on the right. The phase angle for the TAW observations shows highly Alfv\'{e}nic behavior, with the phase angle oscillating close to $\Delta \Omega = \pi = 180^{\circ}$ for most of the time period before dropping off, while the behavior for the blob observations shows no such pattern, indicating a non-Alfv\'{e}nic structure.

%
%

\begin{table*}[htb] 
\begin{center}
\begin{tabular}{|c|c|c|c|c|c|}\hline
Label & $\min{R_W}$ & $\max{R_W}$ & $\langle R_W \rangle \pm \sigma_{R_W}$ & $ \langle \Delta \Omega \rangle \pm \sigma_{\Delta \Omega}$ & Type \\
\hline
\textbf{S1} &  0.09 &  1.51   &  0.56 $\pm$ 0.39 & $152 \pm 32 \;\;^{\circ}$ & wave \\
\textbf{S2} &  0.11 &  1.93   &  0.52 $\pm$ 0.33 & $139 \pm 47 \;\;^{\circ}$ & wave \\
\hline
\textbf{E1a} & 1.71 & 25.3  & 6.9 $\pm$ 5.8 & $45 \pm 44 \;^{\circ}$ & blob \\
\textbf{E1b} & 1.82 & 18.7  & 4.3 $\pm$ 2.9 & $102 \pm 40 \;^{\circ}$ & blob \\
\textbf{E1c} & 0.33 & 26.6  & 3.8 $\pm$ 4.6 & $98 \pm 29 \;^{\circ}$ & blob \\
\textbf{E1d} & 0.30 & 21.9  & 2.3 $\pm$ 3.3 & $110 \pm 23 \;^{\circ}$ & blob \\
\hline
\textbf{E2a} &  0.04 &  16.1 &  1.3 $\pm$ 2.1 &     $110 \pm 44 \;^{\circ}$ & blob \\
\textbf{E2b} &  0.30 &   3.6  &  0.94 $\pm$ 0.71 & $72 \pm 38 \;^{\circ}$ & blob \\
\hline
\end{tabular}
\end{center}
\caption{Event-averaged Wal\'{e}n Numer ($R_W$) and relative phase difference ($\Delta \Omega$) between $\delta \boldsymbol{B_{\perp}}$ and $\delta \boldsymbol{V_{\perp}}$} for the TAW and streamer blob events. More Alfv\'{e}nic features have a Wal\'{e}n number closer to 1 and a relative phase difference closer to 180$^{\circ}$. 
\label{tab:dhdt}
\end{table*}

\section{Origins of Structured Variability in the Slow Solar Wind}
\label{sxn_disc}

\subsection{Streamer Blobs as 3D Magnetic Flux Ropes from Reconnection in the HCS}\label{sxn_disc_hcs}

As seen in Sections~\ref{ssxn_system} and \ref{ssxn_mhdrxn}, our
simulation has streamer blobs throughout the HCS which are all
formed through helmet streamer pinch-off reconnection \citep{Sheeley1997,
WangYM2000}. 
Earlier 2D and 2.5D MHD simulations have also observed
this process 
\citep[e.g.][]{Einaudi1999, Einaudi2001, Endeve2003, Endeve2004, Rappazzo2005, Lapenta2008blob, Allred2015}. 
Our results suggest that the HCS may be a region
of consistently complex topology, with threaded flux ropes creating
a dynamic layer of slow solar wind surrounding the HCS. 

Such a scenario has been presented by \citet{Crooker1996hps,
Crooker2004} who argue that the in situ observations of the HCS and 
plasma sheet represent a tangled network of squashed flux ropes.
Our 3D magnetic islands are lacking the observed axial field
enhancement, $B_\phi$, corresponding
to a guide field component during the reconnection at the streamer
belt cusp. However, our current simulation results are certainly consistent 
with the \citet{Crooker1996hps} scenario, especially if we were to take 
into account the heliospheric evolution of a more realistic HCS shape, 
fast coronal hole wind, and the resulting stream structure and interaction 
regions. Recent results indicate that turbulence can generate a whole distribution of small-flux ropes in the solar wind throughout the heliosphere \citep[e.g][]{Zheng2018}, but our simulation results show that 3D streamer blob pinch-off reconnection \emph{at the Sun} can contribute a significant component of this variability.

While the tangled nature of the flux ropes should be
consistent with the real Sun and heliosphere, the idea that these flux ropes
should all be completely squashed is not. The helicity condensation
picture of \citet{Antiochos2013} predicts that there ought to be a
guide field component from the shear which should be continuously
generated by convective photospheric evolution. This shear is transported
through the closed flux region towards the closed-open flux boundary
of the helmet streamer belt \citep[see also][]{Knizhnik2015,Knizhnik2017}.
A distribution of various guide field component strengths along the
closed-open flux boundary will become a distribution of axial/core
field strengths during the magnetic island flux rope formation.
Therefore, we expect a less-idealized simulation to generate streamer
blob magnetic structures in the slow wind around the HCS that both match the observed,
well-defined in situ small-scale flux ropes with an enhanced core
field as well as flux ropes without core fields that may be more
easily compressed into planar-like structures during their heliospheric
evolution. Our simulation results, combined with the in situ observations 
of un-squashed flux ropes \citep[as argued by][]{Moldwin2000}, suggest 
that both may be present in the HSC-associated slow solar wind, essentially all of the time.

One of the major implications of a highly structured heliospheric
plasma sheet filled with magnetic islands is related to energetic
particle acceleration in the heliosphere. The formation and evolution
of magnetic island structures have been suggested as processes for
accelerating particles in and around magnetic reconnection sites
\citep[e.g.][and references therein]{Drake2006b, Dahlin2014,
Guidoni2016, Khabarova2016} where particles can be accelerated via
the curvature drift and Fermi-reflection associated with the
contraction of the island flux surfaces.

Our particular situation
in which we have many dynamic islands with potentially small guide
fields is an interesting one to consider. \citeauthor{Dahlin2014}
(\citeyear{Dahlin2014}, \citeyear{Dahlin2015}, \citeyear{Dahlin2016})
used particle-in-cell simulations to show that when the guide field is
much larger than the reconnecting field then electron acceleration
via the Fermi mechanism is suppressed. On the other hand, turbulent 3D simulations show that with no guide field at all electron
acceleration doesn't occur either. Too little guide field
and particle acceleration is negligible; too much guide field and
not only is the particle lost from the island structure quickly,
but there is less flux surface contraction, limiting the contribution
of the second mechanism.

A complex network of streamer-blob  flux ropes in
the heliospheric plasma sheet with a range of guide fields 
is an ideal breeding ground for particle acceleration. Indeed, 
\citeauthor{Khabarova2015} (\citeyear{Khabarova2015},
\citeyear{Khabarova2016}) have performed calculations
of the energization due to magnetic islands to explain the in situ
observations of keV--MeV particles in the HCS and discussed the
importance of this suprathermal component to generating the observed
intensities of CIR and CME shock-associated SEPs.

\subsection{Torsional \Alfven Waves as ``Pseudo-Flux Rope'' Structures from Interchange Reconnection}

The TAW in our simulation, described in Section~\ref{ssxn_mhdwave},
was generated by a large-scale rotational motion on the solar
surface. Because this rotational motion was much larger and more coherent
than any photospheric motion on the real Sun, it produced an isolated, large-scale 
TAW signature which was easily investigated. 
While there are many observations of TAWs in the solar wind
(\citealt{Yu2016}; also the STEREO PLASTIC Level 3 data 
product\footnote{\url{https://stereo-ssc.nascom.nasa.gov/data/ins_data/plastic/level3/Alfven_Waves/}}) 
there is some inherent uncertainty associated with both visual 
inspection and automated identification and classification of 
small-scale flux ropes and TAWs \citep[e.g.][]{Feng2010comment, 
Cartwright2010reply, Yu2014}. There is even a documented case 
of a TAW \textit{within} a small-scale flux rope \citep{Gosling2010taw}. Here, we were able to analyze an oversized but clean sample 
TAW and compare it with HCS-associated structures under the same conditions.

One proposed mechanism for TAW creation at the Sun includes the reconnection 
associated with solar jets \citep{Shibata1986}. Explosive jet observations 
provide some of the most direct measurements of the \Alfvenic propagation 
of magnetic field line twist into the solar corona and into the heliosphere 
\citep[see][and references therein]{Raouafi2016, Uritsky2017}. The high-resolution 
simulations by \citet{Wyper2016b} resolved \emph{both} the large-scale 
twist jet eruption along the external spine line \emph{and} small-scale, 3D flux rope islands with localized, concentrated twist which were formed
in the reconnection current sheet layer during the breakout/interchange
reconnection prior to the main eruption \citep[see
also][]{Lynch2014,Wyper2017}. \citet{Wyper2016b} described the
propagation of the magnetic island's localized twist along the newly
reconnected field lines as TAWs.
Similar interchange reconnection dynamics are also expected over a 
distribution of larger spatial scales from active region flux systems \citep{Torok2009,Archontis2013,MorenoInsertis2013,Lynch2014} to
coronal pseudostreamers \citep{Torok2011, Zuccarello2012a, Masson2014,
Edmondson2017, WangYM2018}. 

While most of these simulations showing TAW formation are driven by either emerging twisted
flux or twisting up existing flux distributions (i.e. introducing
some form of twist that is ultimately transferred via reconnection
onto open field lines), there is increasing evidence that sufficiently
resolved reconnection processes---even in planar geometries---will
introduce a twist component during the formation and ejection of
3D magnetic island plasmoids. 
For example, the simulations of \citet{Edmondson2017} investigated 
the generation of 3D magnetic island flux ropes from \emph{externally}-driven 
interchange reconnection in a coronal pseudostreamer geometry.
These small-scale flux rope structures had a significant localized twist 
component even though the simulation boundary flows were purely 
translational (i.e. no twist component). 
As the islands were ejected within the reconnection exhaust into the open 
field region, their core/axis tended to realign {towards the} open field 
direction so that the localized twist component was able to more freely 
propagate away as ``pseudo-flux rope'' TAWs.
\citet{Edmondson2017} argued this evolution represents the small-scale, 
reconnection-generated 3D magnetic island manifestation of the larger-scale 
\citet{Shibata1986} twist-jet scenario.

Since a wide variety of reconnection scenarios can generate 
these ``pseudo flux rope" and flux rope signatures and the observed variability and composition 
of the slow solar wind strongly suggests its origin is related to 
magnetic reconnection \citep{Zurbuchen2007,ZhaoL2017}, we conjecture that these periods of structured and 
coherent variability in the slow solar wind should also exhibit 
temporal coincidence with other plasma and composition 
signatures of reconnection. Encouragingly, there are observational 
studies show that this may be the case \citep{Feng2015a,Feng2015b,Yu2016,Kepko2016, WangYM2018}. These TAWs, generated by interchange reconnection in the solar corona, should therefore be observed in the S-Web-associated slow solar wind discussed by \citet{Antiochos2011}, \citet{Higginson2017a} and \citetalias{Higginson2017b}, both along S-Web arcs and in the HCS.

\section{Conclusions}
\label{sxn_conc}

We have reported on two distinct features of structured slow 
solar wind variability present in the \citetalias{Higginson2017b} 
MHD simulation. 
First, we presented the analysis of the basic plasma and field 
structure of 3D, streamer-blob flux ropes created in the HCS by pinch-off 
reconnection at the helmet streamer cusp, which  
represents a critical missing component of solar wind modeling thus far. This form of solar wind variability must be included in 
future models in order to correctly simulate true heliospheric observations. Steady-state MHD 
models by their very nature can not include helmet streamer dynamics, which are observed both 
remotely and in-situ. Our dynamic simulation results are consistent with many aspects of the in situ 
observations of the small-scale magnetic flux ropes {and} suggest 
that even the simplest, equatorial heliospheric plasma sheet 
could be filled with tangled and intermingled flux rope structures, 
as in the \citet{Crooker1996hps} picture, making it a promising region 
for particle acceleration. Additionally, we {discussed} the lack of guide field in our helmet streamer and the flattened flux ropes we observed in the HCS. We predict that the guide field component of observed HCS flux ropes will be a direct measure of the helicity condensation rate at the open magnetic field boundary. 

Second, we analyzed an idealized, large-scale TAW that propagates 
along the S-Web arc in the slow solar wind to high latitudes. We examined the similarities and 
differences of {these} \Alfvenic field and plasma fluctuations to those in the 
HCS streamer-blob  flux ropes and discussed interchange reconnection 
scenarios that are likely to result in generation of TAWs in the corona. These TAWs should be found in the slow solar wind all along the S-Web, both in S-Web arcs and the HCS, where interchange reconnection should be prevalent \citepalias{Higginson2017b}. This type of 
variability observed in the slow solar wind is most likely formed in principio, and may hold important clues as to the different sources of different types of wind. Our simulation results and analysis are an important {contribution} towards 
characterizing the field and in situ plasma signatures predicted by the 
\citet{Antiochos2011} dynamic S-Web model 
in anticipation for the Parker Solar Probe and Solar Orbiter missions. 


\acknowledgments

A.K.H. acknowledges support from the NASA LWS TR\&T and HSR programs. 
B.J.L. acknowledges support from NASA NNX15AJ66G and NNX15AB69G. 
The authors thank S.~K.~Antiochos, C.~R.~DeVore, and the rest of the GSFC 
Team Blobbers (N.~Viall, L.~Kepko, J.~Allred, M.~Schlenker, and P.~MacNeice) for 
extremely helpful discussion and comments during preparation of the manuscript.

%



\bibliographystyle{apj} 
\bibliography{smfr}  

\end{document}